\documentclass[11pt,a4paper]{article}
\usepackage{jheppub}
\usepackage{amssymb}
\usepackage{mathrsfs, amsmath}
\usepackage{fancyhdr}
\usepackage{xcolor}
\usepackage{ifpdf}
\usepackage{rotating}
\usepackage{comment,tikz}
\usepackage{url}
\definecolor{darkblue}{rgb}{0,0,0.6}
\usetikzlibrary{patterns}

\def\D{\Delta}

\newcommand{\tf}{\tilde{f}(\y)}

\newcommand{\va}{\textrm{vac}}

\newcommand{\f}{\frac}

 \def\p{\partial}

\newcommand{\EV}{E_{\text{vac}}}
\newcommand{\y}{\mathbf{y}}
\newcommand{\Mi}{{\mathcal{M}_i}}
\newcommand{\Md}{{\mathcal{M}_{d-1}}}
\newcommand{\Mz}{{\mathcal{M}_{0}}}

\newcommand{\evac}{\varepsilon_{\text{vac}}}

\newcommand\rref[1]{(\ref{#1})}

\newlength{\slength}
\def\e{{\epsilon}}
\def\p{\partial}

\def\a{\alpha}
\def\b{\beta}

\newcommand{\bi}{\begin{itemize}}
\newcommand{\ei}{\end{itemize}}
\newcommand{\bea}{\begin{eqnarray}}
\newcommand{\eea}{\end{eqnarray}}
\newcommand{\be}{\begin{equation}}
\newcommand{\ee}{\end{equation}}

\newcommand*\holland{\includegraphics[scale=0.02]{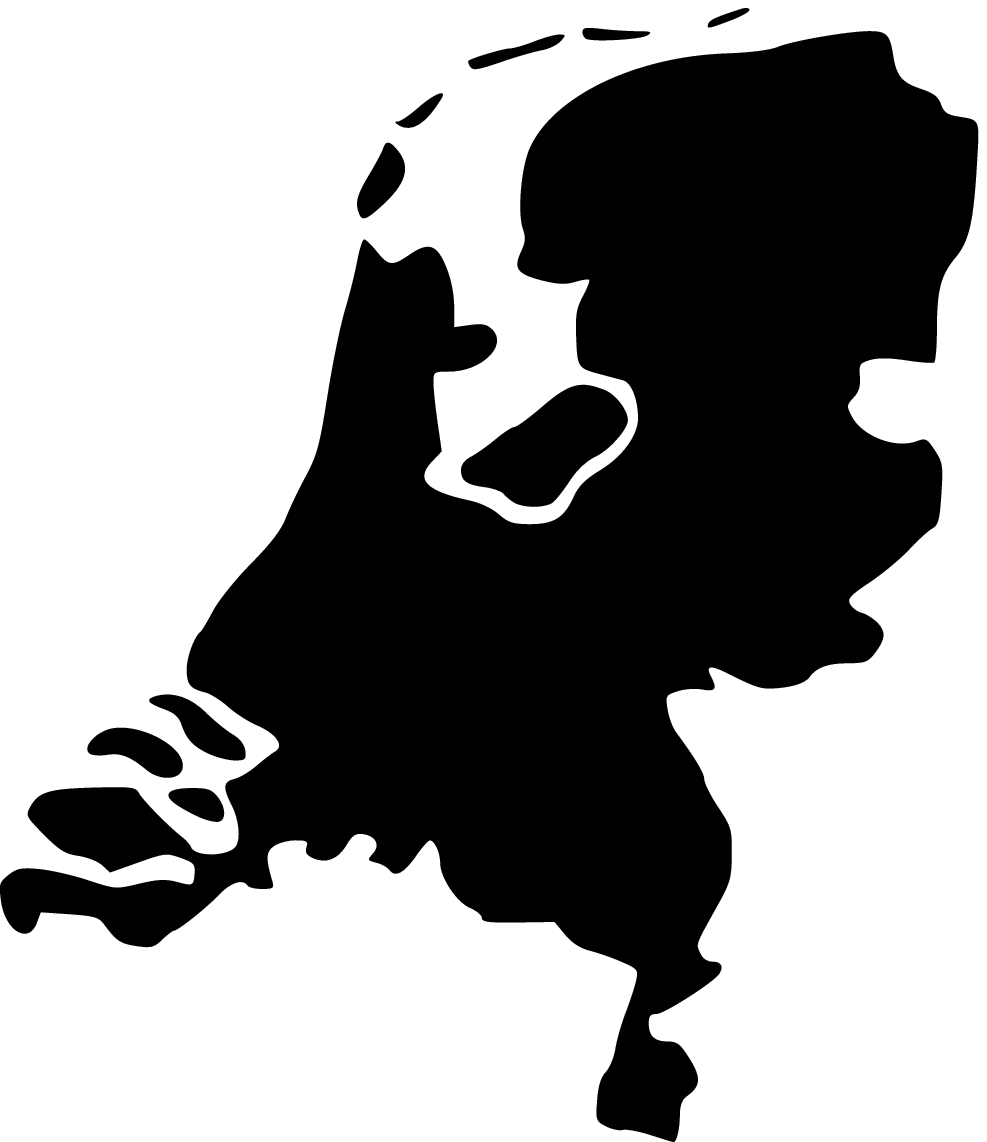}}
\newcommand*\cali{\includegraphics[scale=0.07]{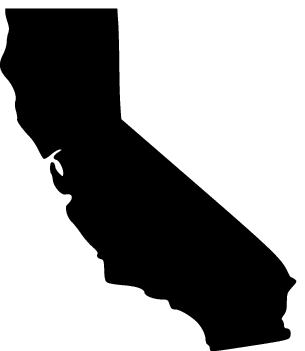}}

\DeclareRobustCommand{\cube}[3]{
\vcenter{\hbox{\tikz[baseline=15pt,scale=0.6,line width = 0.5pt]{
  \draw (0.6,-0.5,0) coordinate (x) |- (-0.4,0.5,0) coordinate [midway] (h) coordinate (y) -- (-0.4,0.5,1) coordinate (a) -- (-0.4,-0.5,1) coordinate (z) -- (0.6,-0.5,1) edge (x) -- (0.6,0.5,1) coordinate (v) edge (h)
  -- (a)  ;
  \node at (-0.8,0,1) {$ #1 $};
  \node at (0.1,-0.9,1) {$ #2 $};
  \node at (1,-0.7,0.75) {$ #3 $};
}}}}

\numberwithin{equation}{section}
\allowdisplaybreaks  

\title{Universality of sparse $d>2$ conformal field theory at large $N$}

\author[\holland]{Alexandre Belin,}
\author[\holland]{Jan de Boer,}
\author[\holland]{Jorrit Kruthoff,}
\author[\cali]{Ben Michel,}
\author[\cali]{Edgar Shaghoulian,}
\author[\cali]{and Milind Shyani}

\affiliation[\holland]{Institute for Theoretical Physics Amsterdam and Delta Institute for Theoretical Physics, University of Amsterdam, Science Park 904, 1098 XH Amsterdam, The Netherlands}
\affiliation[\cali]{Department of Physics, University of California, Santa Barbara, CA 93106}

\abstract{We derive necessary and sufficient conditions for large $N$ conformal field theories to have a universal free energy and an extended range of validity of the higher-dimensional Cardy formula. These constraints are much tighter than in two dimensions and must be satisfied by any conformal field theory dual to Einstein gravity. We construct and analyze symmetric product orbifold theories on $\mathbb{T}^{d}$ and show that they only realize the necessary phase structure and extended range of validity if the seed theory is assumed to have a universal vacuum energy.}

\keywords{CFT, holography, AdS/CFT}

\arxivnumber{1610.06186}

\begin{document}

\maketitle

\newpage

\setcounter{page}{1}
\pagenumbering{arabic}

\setcounter{tocdepth}{2}



\section{Introduction}
The strongest form of the AdS/CFT correspondence states that every conformal field theory (CFT$_d$) is dual to a theory of quantum gravity living in a higher-dimensional anti-de Sitter space  (AdS$_{d+1}$). For a generic CFT, the dual theory of quantum gravity at low energies will look nothing like semi-classical Einstein gravity. One of the most interesting questions in the context of holography is then to understand which CFTs -- when interpreted as theories of quantum gravity in AdS -- have a semi-classical Einstein gravity limit.

The most straightforward constraint emerging from the AdS/CFT dictionary for a semi-classical bulk is that the CFT should have a large number of degrees of freedom, usually parameterized by $N$. Large $N$ in the field theory implies a semi-classical bulk since its inverse scales as a positive power of the Planck length in AdS units: $N^{-1}\sim (\ell_P/\ell_{\textrm{AdS}})^{\#}$ for $\#>0$. This is the bulk expansion parameter controlling AdS-scale quantum gravitational effects.\footnote{To have a theory that looks like Einstein gravity at low energies, we also need an expansion parameter that can suppress higher-spin fields. The 't Hooft coupling in gauge theory usually plays the role of this expansion parameter. Interestingly, like in the D1-D5 duality, certain features of Einstein gravity can be reproduced without explicitly invoking this assumption. We will not explicitly implement any constraints on our field theories with the purpose of decoupling higher-spin fields.}

Besides large $N$, a semi-classical theory of gravity in anti-de Sitter space has many universal features that must be encoded in any putative dual CFT. To explore the emergence of gravity from field-theoretic degrees of freedom, it is natural to try to reproduce these universal features by implementing some additional assumptions on a generic large-$N$ CFT. There has been tremendous progress in this direction for the case of three-dimensional gravity \cite{Fitzpatrick:2015zha, Fitzpatrick:2015foa, Fitzpatrick:2015dlt, Fitzpatrick:2016thx, Fitzpatrick:2016ive, Chen:2016cms, Fitzpatrick:2016mjq, Roberts:2014ifa, Hartman:2013mia, Hartman:2014oaa, Asplund:2014coa, Asplund:2015eha, Perlmutter:2016pkf, Anous:2016kss}, throughout which large central charge and a sparse low-energy spectrum play a prominent role. These powerful methods for the most part rely on the fact that all stress tensor interactions in the CFT are captured by the Virasoro block of the identity, which is assumed to dominate. The success of this particular approach is related to the topological nature of gravity in three dimensions, which precludes obvious generalizations to higher dimensions. Nevertheless, it is a compelling problem to reproduce features of higher-dimensional AdS gravity purely from the CFT. A small sample of work in this direction includes \cite{Kovtun:2008kw, Heemskerk:2009pn, ElShowk:2011ag, Komargodski:2012ek,  Fitzpatrick:2012yx, Fitzpatrick:2012cg, Fitzpatrick:2014vua, Fitzpatrick:2015qma, Camanho:2014apa, Hartman:2015lfa, Maldacena:2015iua, Alday:2014tsa}.

In this paper, we will focus on a technical tool that has received little exposure in higher dimensions: modular invariance. For 2d CFTs, modular invariance can be used to precisely determine how sparse the spectrum should be to reproduce the thermal phase structure of 3d gravity \cite{Hartman:2014oaa} (see \cite{Benjamin:2015hsa} for a similar consideration in supersymmetric theories). For theories obeying this sparseness constraint, the Cardy formula \cite{Cardy:1986ie} -- which is usually only valid asymptotically  as $\Delta/c\rightarrow \infty$ -- has an extended regime of validity down to energies $\Delta \sim c$. This precisely matches the bulk phase structure since the black holes begin dominating the ensemble at $\Delta \sim c$.

The relevance of modular invariance in higher-dimensional holographic CFTs has been much less explored. In \cite{Shaghoulian:2015kta,  Shaghoulian:2015lcn}, it was shown that modular invariance of the torus partition function implies the existence of an asymptotic formula that correctly reproduces the Bekenstein-Hawking entropy of the dual black brane. This formula is the higher-dimensional generalization of the Cardy formula and only holds in the limit of large energy for generic CFTs. Holographic CFTs, on the other hand, must have an extended range of validity of this formula as implied by the bulk phase structure. The goal of this paper is to further exploit modular invariance and place constraints on CFTs such that they have this extended range of validity. We also want to match the precise phase structure of gravity, which is much richer than in two dimensions and exhibits both quantum and thermal phase transitions. One of the key challenges that we will face is that the functional form of the vacuum energy in higher dimensions is not uniquely fixed by conformal invariance, although we will discover several nontrivial constraints due to modular invariance.

We can summarize our results as follows. A general CFT on $\mathbb{T}^d$ will have an extended Cardy formula and a universal phase structure if and only if 
the partition function is dominated by the vacuum contribution when quantizing along any cycle but the shortest one. Proving this will require using the modular constraints on the vacuum energy alluded to above. From here, we will consider large-$N$ theories and exhibit distinct sets of necessary and sufficient sparseness conditions on the spectrum to achieve this vacuum domination. 


In analyzing calculable theories that satisfy these necessary and sufficient conditions, and which therefore have a universal free energy, we are led to the construction of symmetric orbifold theories in higher dimensions. Symmetric orbifolds have been analyzed in great depth in two dimensions \cite{Dixon:1986qv, Dijkgraaf:1996xw, Bantay:1997ek, Bantay:2000eq, Lunin:2000yv,  Pakman:2009zz, Keller:2011xi}, and play an explicit role in the D1-D5 duality \cite{deBoer:1998kjm, Dijkgraaf:1998gf, Seiberg:1999xz}. Still, they have not explicitly appeared in holographic dualities in higher dimensions nor, to the best of our knowledge, have they been constructed. For their construction, we use a similar procedure as in two dimensions to build a modular invariant partition function. This includes both untwisted and twisted sectors. For large-$N$ symmetric product orbifolds, the density of states of the untwisted sector is shown to be slightly sub-Hagedorn, whereas for the twisted sector it is precisely Hagedorn. Saturation of the necessary and sufficient conditions for universality is then guaranteed by assuming that the subextensive parts of the vacuum energy vanish. This assumption constrains the choice of seed theory we can pick. This is somewhat of a loss of generality compared to two dimensions but can be expected by the increasing richness of CFTs in higher dimensions. Provided we pick the seed accordingly, the symmetric orbifolds reproduce the phase structure of higher-dimensional AdS gravity: they have an extended regime of validity of the Cardy formula and a Hagedorn transition at precisely the same temperature as the Hawking-Page transition in the bulk. \\
\\
The paper is organized as follows. We start in section \ref{generalities} with a general discussion of CFTs on $d$-dimensional tori and modular invariance.  In section \ref{phases} we summarize the phase structure of toroidally compactified gravity in anti-de Sitter spacetime. These two sections set the stage for the meat of the paper. Section \ref{proofs} is dedicated to a detailed discussion of the necessary and sufficient conditions that are required to have a universal free energy. The implementation of these conditions is then explored in section \ref{section:orbifolds}. We discuss the construction of orbifold theories on $d$-dimensional tori and show that symmetric product orbifolds have a universal free energy. We conclude with a discussion and outlook in section \ref{disc}. The appendices contain additional material, including extensions to the case with angular momentum and calculations translating the results from the canonical partition function to the microcanonical density of states.

\section{Generalities of CFT$_{d}$}\label{generalities}
We now introduce some of the basic technology of modular invariance that we will use to derive our general CFT results. For more details see \cite{Shaghoulian:2015kta, Shaghoulian:2015lcn}. In this paper we will study conformal field theories defined on a Euclidean $d$-torus $\mathbb{T}^d$. We fix the metric on this torus to be
\be
ds^2 = dx_0^2 + dx_1^2 + \dots + dx_{d-1}^2
\ee
with identifications
\be
(x_0,x_1,..,x_{d-1}) \sim (x_0,x_1,..,x_{d-1}) + \sum_{i=0}^{d-1} n_i U_i \,.
\ee
where $U_i$ are vectors defining the torus $\mathbb{T}^d$ and the $n_i$ are integers. These vectors can be conveniently organized in a matrix as 
\begin{equation}
U = (U_0 \ \ \cdots \ \ U_{d-1})^T = \begin{pmatrix}\label{Umatrix}
L_0 & \theta_{01}   & \cdots   &\theta_{0,(d-2)}& \theta_{0,(d-1)}  \\
0  & L_1    & \cdots &\theta_{1,(d-2)}& \theta_{1,(d-1)} \\
\vdots  & \vdots & \ddots   & \vdots& \vdots\\
0  &  0 &\cdots     &L_{d-2} & \theta_{(d-2),(d-1)}  \\
0  &  0 & \cdots     &0&L_{d-1}
\end{pmatrix}
\end{equation}
and define a $d$-dimensional lattice of identifications. This matrix contains the lengths of the cycles along its diagonal and the $\theta_{ij}$ capture all possible twists of the torus $\mathbb{T}^d$. Modular invariance of the torus partition function for conformal field theories is a powerful constraint on the theory. The invariance can be stated as the action of large conformal transformations on the lattice spanned by the set $\{U_i\}$. These large conformal transformations form the group $SL(d,\mathbb{Z})$ and act on the matrix $U$ in \rref{Umatrix} by left multiplication. $SL(d,\mathbb{Z})$ is generated by two elements \cite{Trott:1962}
\be \label{sldgen}
S = \begin{pmatrix}
0 & 1 & 0 & \dots & 0 & 0\\
0 & 0 & 1 & \dots & 0 & 0\\
\vdots & \vdots & \vdots & \ddots & \vdots & \vdots \\
0 & 0 & 0 & \dots & 0 & 1\\
(-1)^{d+1} & 0 & 0 & \dots & 0 & 0
\end{pmatrix},\quad
T = \begin{pmatrix}
1 & 1 & 0 &  \dots & 0 & 0\\
0 & 1  &0 &  \dots & 0 & 0\\
\vdots & \vdots & \vdots & \ddots & \vdots & \vdots \\
0 & 0 & 0 & \dots & 1 & 0 \\
0 & 0 & 0 &  \dots & 0 & 1
\end{pmatrix}\,.
\ee 
They can be shown to generate any pairwise swap and a twist along any direction. 
For even $d$, we quotient by the center of the group $\{-1,1\}$ to obtain $PSL(d,\mathbb{Z})$, but for simplicity we will universally refer to the group as $SL(d,\mathbb{Z})$. Using scale invariance to unit-normalize one of the cycle lengths shows that we have $(d-1)(d+2)/2$ real moduli captured by the matrix $U$. 

In spacetime dimension greater than two, modular transformations generically change the spatial background of the theory (i.e. change the Hilbert space), making it difficult to relate the low-lying states to the high-lying states on a fixed background. However, as discussed in \cite{Shaghoulian:2015kta} there exist two choices of torus which allow for a high-temperature/low-temperature duality to be considered. The first is the background $S^1_\beta \times S^1_L \times \mathbb{T}_{L_\infty}^{d-2}$, where $L_\infty\gg \beta, L, \beta^2/L$. In this case by appealing to extensivity in the large directions we have the approximate invariance 
\be
\log Z(\beta) \approx (L/\beta)^{d-2} \log Z(L^2/\beta)\,.
\ee
This can be transformed into an exact high-temperature/low-temperature duality by passing to a density defined by dividing $\log Z(\beta)$ by the volume of the large torus as it decompactifies, but we will not pursue that here. 

To produce an exact invariance on a compact manifold, we can also consider a special torus given by $S^1_\beta \times S^1_L \times S^1_{L^2/\beta}\times \cdots \times S^1_{L^{d-1}/\beta^{d-2}}$, for which
\be
Z(\beta) = Z(L^d/\beta^{d-1})\,.
\ee
This invariance is obtained by an $SL(d,\mathbb{Z})$ transformation and a scale transformation. It will play an important role in our CFT analysis.

To deal with the case of a general torus where there is no high-temperature/low-temperature duality, we will find it useful to define some notation. For a $d$-dimensional torus of side lengths $L_0, L_1, \dots, L_{d-1}$, where $\beta = L_0$, we will denote the partition function quantized in an arbitrary channel as:
\be
Z[\mathcal{M}^d] = Z(L_i)_{\mathcal{M}_i} = \sum e^{-L_i E_{\mathcal{M}_i}}\,.
\ee
$Z[\mathcal{M}^d]$ denotes the Euclidean path-integral representation of our partition function, which treats space and time democratically. The next form of the partition function picks direction $i$ as time and gives a Hilbert space interpretation of the path integral. Since the spatial manifold will change depending on which direction is chosen as time, we use the notation $\mathcal{M}_i$ to explicitly denote the spatial manifold. It is defined as $\mathcal{M}^d = \mathcal{M}_i \times S^1_{L_i}$. Brackets will always imply a Euclidean path-integral representation while parentheses will imply a Hilbert-space representation.

\subsection{Review of higher-dimensional Cardy formulas}
Now we will provide a derivation of the higher-dimensional Cardy formula on an arbitrary spatial manifold $ S^1_\beta \times X$. We will only need the result for a spatial torus, but we will keep the discussion general. The fact that modular transformations generically change the Hilbert space of the torus partition function will not provide an obstruction, although we will see in the resulting formulas that our high-temperature partition function and asymptotic density of states refer to the vacuum energy on a \emph{different} spatial background in general.

 We assume our theory to be local, modular invariant, and to have a spectrum of real energies on the torus that is bounded below by an energy that is discretely gapped from the rest of the spectrum. At asymptotically high temperature $\beta/V_X^{1/(d-1)}\rightarrow 0$, we can use extensivity of the free energy to replace our spatial manifold $X$ with a torus $\mathbb{T}^{d-1}$ of cycle lengths $L_1\leq L_2\leq \cdots \leq  L_{d-1}$ and no twists, with $V_X=L_1\cdots L_{d-1}\equiv V_{\mathcal{M}_0}$. We therefore have
\be\label{ext}
Z[S^1_\beta \times X] = Z(\beta)_X\approx Z(\beta)_{\mathcal{M}_{0}}= \sum e^{-\beta E_{\mathcal{M}_0}} \approx e^{\tilde{c} V_{\mathcal{M}_0}/\beta^{d-1}}
\ee
at asymptotically small $\beta$ for some thermal coefficient $\tilde{c}>0$. This thermal coefficient is not a priori related to any anomalies except in two dimensions. Considering a quantization along $L_{d-1}$ gives us
\be
Z(L_{d-1})_{\mathcal{M}_{d-1}} = \sum e^{-L_{d-1} E_{\mathcal{M}_{d-1}}} = e^{-L_{d-1} E_{\va, \mathcal{M}_{d-1}}} \sum e^{-L_{d-1} (E-E_\va)_{\mathcal{M}_{d-1}}}\,.
\ee
For $d=2$ in a scale-invariant theory, $\beta$ becoming asymptotically small is equivalent to $L_{d-1}$ becoming asymptotically large, since only the ratio $L_{d-1}/\beta$ is meaningful. However, for $d>2$ we have the additional directions $L_i$ which may prevent us from interpreting the quantization in the $L_{d-1}$ channel as a low-temperature partition function which projects to the vacuum. To deal with this, consider the limit $L_{d-1}\rightarrow \infty$ where we indeed project efficiently to the vacuum:
\be
\lim_{L_{d-1}\rightarrow \infty} \f{\log Z(L_{d-1})_{\mathcal{M}_{d-1}}}{L_{d-1}} = - E_{\va, \mathcal{M}_{d-1}}\,.
\ee 
Using $Z(\beta)_{\mathcal{M}_{0}} = Z(L_{d-1})_{\mathcal{M}_{d-1}}$ gives us $E_{\va,\mathcal{M}_{d-1}} = -\tilde{c} V_{\mathcal{M}_{d-1}}/\beta^{d}$.  We are therefore able to extract the scaling of the vacuum energy as $E_{\va, \mathcal{M}_{d-1}} \propto - V_{\mathcal{M}_{d-1}}/\beta^d$ as $\beta\to0$. The proportionality coefficient, which we define as $\varepsilon_\va$, is $\varepsilon_\va = \tilde{c}$. Furthermore, notice that $E_{\va, \mathcal{M}_{d-1}}$ is clearly independent of $L_{d-1}$, so this result is general even though we took the limit $L_{d-1}\rightarrow \infty$ to obtain it. In the general case of arbitrary $L_{d-1}$ we can therefore write for  $\beta\to0$
\be
Z(L_{d-1})_{\mathcal{M}_{d-1}} = e^{\tilde{c} V_{\mathcal{M}_{0}}/\beta^{d-1}} \sum e^{-L_{d-1} (E-E_{\va})_{\mathcal{M}_{d-1}}}\,.
\ee
Again equating with $Z(\beta)_{\mathcal{M}_0}$, we see that the excited states must contribute at subleading order, since the vacuum contribution is sufficient to obtain $Z(L_{d-1})_{\mathcal{M}_{d-1}} = Z(\beta)_{\mathcal{M}_{0}}$ at leading order in small $\beta$. The concern over the directions $L_i$ and poor projection to the vacuum alluded to earlier is therefore not a problem at leading order. We are finally left with 
\be
S(\beta) = (1-\beta \p_{\beta})\log Z(\beta)_X \approx dV_X \varepsilon_\va/\beta^{d-1}
\ee
for the high-temperature entropy of a modular-invariant CFT on an arbitrary spatial background $X$. 

Now we consider the implications for the density of states:
\begin{align}
\rho(E_s) &=\f{1}{2\pi i} \int_{\alpha-i\infty}^{\alpha+i\infty} d\beta \,Z(\beta)_X e^{\beta E_s}\\
&= \f{1}{2\pi i} \int_{\alpha-i\infty}^{\alpha+i\infty} d\beta \left(e^{- \varepsilon_\va V_X /\beta^{d-1}}\sum e^{-\beta E }\right) e^{\varepsilon_\va V_X /\beta^{d-1} + \beta E_s}\,,
\end{align}
for some $\alpha>0$. Performing a saddle-point on the part of the integrand outside of the parentheses and evaluating the integrand on this saddle $\beta_s \propto E_s^{-1/d}$ gives us the higher-dimensional Cardy formula:
\be
\log \rho(E_s) = \f{d}{(d-1)^{\f{d-1}{d}}} (\varepsilon_\va V_X)^{\f{1}{d}} E_s^{\f{d-1}{d}}\,. \label{cardyhigherd}
\ee
 The saddle point implies $\beta_s \rightarrow 0$ as $E_s\rightarrow \infty$. To ensure that this saddle point is valid, we need to check that the part of the integrand in the parentheses, which we call $\tilde{Z}_X(\beta)$, does not give a big contribution on the saddle:
\be
\tilde{Z}_X(\beta_s) = e^{- \varepsilon_\va V_X /\beta_s^{d-1}}\sum e^{-\beta_s E} \,.
\ee
From high-temperature ($\beta_s  \rightarrow 0$) extensivity \eqref{ext}, we know that we can write this as 
\be
\tilde{Z}_X(\beta_s) \approx e^{-\varepsilon_\va V_X/\beta_s^{d-1}} e^{\tilde{c} V_X/\beta_s^{d-1}}= 1\,,
\ee
where we used $\tilde{c}=\varepsilon_\va$ (and one notices $\tilde{c}$ is independent of spatial background by replacing the high-temperature partition function on the given manifold with the high-temperature partition function on a torus of spatial lengths $L_1,\dots L_{d-1}$ with $V_{\mathcal{M}_0}=V_X$). Our saddle-point approximation is therefore justified, and we have the higher-dimensional Cardy formula as advertised. 

In particular, considering the spatial background to be $X=S^{d-1}$ gives the asymptotic density of local operators by the state-operator correspondence. In the rest of this paper we will only be interested in the CFT on $\mathbb{T}^d$.

\subsection{Review of vacuum energies in CFT}\label{vacreview}
\subsubsection{Normalization of vacuum energy}
In a generic field theory, one is always free to shift the Hamiltonian by an arbitrary constant. This therefore shifts what we call the vacuum energy. Indeed, the well-known Casimir effect demonstrates that derivatives with respect to spatial directions $d E_\va/dL_i$ are the physical observables, leaving an ambiguity in the normalization of $E_\va$. Additional structure, such as supersymmetry or modular invariance, disallows such an ambiguity. Even in a purely scale-invariant theory one can fix the normalization of the vacuum energy. Scale invariance requires that energies, and in particular the ground state energy, scale
as inverse lengths under a rescaling of the spatial manifold: $E_\va(\lambda L_1,\lambda L_2,\ldots)=\lambda^{-1} E_\va(L_1,L_2,\ldots)$. This fixes the shift ambiguity in $E_\va$.

\subsubsection{Subextensive corrections to the vacuum energy}
\begin{figure}
\centering
\includegraphics[scale=0.45]{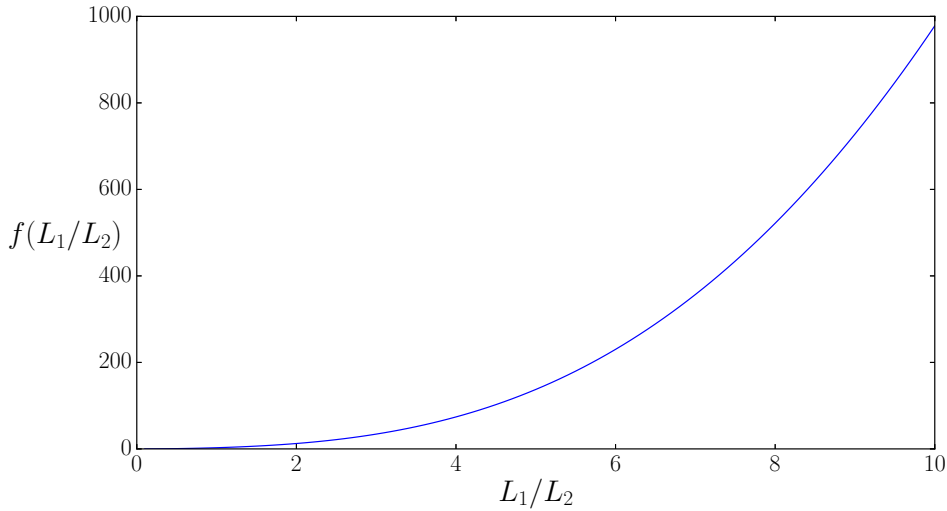}
\caption{\label{bosonCasimir}The functional form of $f(L_1/L_2)$ in the vacuum energy (defined in \eqref{casimirf}) of a free boson in $2+1$ dimensions on a two-torus $\mathbb{T}^2$ with sides $L_1$ and $L_2$. As can be seen in the plot, $f(L_1/L_2)$ is positive and monotonically increasing.}
\end{figure}

The higher-dimensional Cardy formulas involves the vacuum energy density on $S^1 \times \mathbb{R}^{d-2}$, which by its relation to the extensive free energy density in a different channel is negative and has a fixed functional form. If we compactify more directions and make them comparable to the size of the original $S^1$, then we will in general get corrections to the asymptotic formula. For two-dimensional CFT there is only one spatial cycle so no such corrections can enter. To capture the essence of what happens, let us consider a three-dimensional CFT on $S^1_{\beta} \times S^1_{L_1} \times S^1_{L_2}$ with $L_1<L_2$. The low-temperature partition function will project to the vacuum state on $S^1_{L_1} \times S^1_{L_2}$, which can be parameterized as 
\be\label{casimirf}
E_{\va, L_1\times L_2} = -\f{\varepsilon_\va L_2}{L_1^2}(1+f(L_1/L_2))\,.
\ee
Let us define $y=L_1/L_2$. The function $f(y)$ is capturing all of the corrections beyond the asymptotic formula, so we have $f(0) = 0$ and $f(y\rightarrow \infty) = -1+ y^3$. In general, $f(y)$ is a nontrivial function of $y$. Later in the text we will derive some positivity and monotonicity constraints on $f(y)$ by using modular invariance, but for now let us exhibit its functional form for the free boson theory, shown in figure \ref{bosonCasimir}.

In higher dimensions, there are more independent ratios that can be varied, and in general the corrections beyond the asymptotic formula are given by some nontrivial function of $d-2$ dimensionless ratios $y_i = L_1/L_i$ which for simplicity we will often write as $f(\mathbf{y})$ with $\mathbf{y} = (y_2,y_3,\dots, y_{d-1})$.

We will also find it useful to consider the parameterization of the vacuum energy in arbitrary dimension as 
\be
\EV= -\f{\evac V_{d-1}}{L_{\textrm{min}}^d}\left(1+\tilde{f}(\y)\right).
\ee
which always has the smallest cycle in the denominator. The key difference between $\tilde{f}(\y)$ and $f(\y)$ is that it is possible for $\tilde{f}(\y)$ to be identically zero for all values of its arguments, whereas this is not the case for $f(\y)$ as discussed in three dimensions above. We will find, for example, that gravity implies a vacuum energy structure with $\tilde{f}(\y)=0$ up to $1/N$ corrections. We will often just write $\tf=0$, by which we mean the equality up to $1/N$ corrections.


\section{Phase structure of toroidally compactified AdS gravity}\label{phases}
In this section we will recap what is known about the phase structure of gravity in AdS with a toroidally compactified boundary. This phase structure is easy to deduce for pure gravity without spontaneous breaking of translation invariance, which is the case we will restrict ourselves to. The most remarkable feature of this phase structure is the absence of any nontrivial finite-size corrections to the vacuum energy and free energy, up to sharp phase transitions as circles become comparably sized. In other words, the function $\tilde{f}(\y)$ defined in the previous section \emph{vanishes} for all values of its arguments. As usual there will be nonzero contributions suppressed by $1/N$. Note that weakly coupled theories, including e.g. $\mathcal{N}=4$ super Yang-Mills, do not realize this sort of structure \cite{Myers:1999psa}. We will not consider the possibility that the singular solutions used in \cite{Myers:1999psa} are relevant for the phase structure. An argument against them is as follows. Assume that such a singular solution provides the vacuum energy of the theory under multiple compactifications. By the higher-dimensional Cardy formula, there must therefore exist a black brane with higher entropy than AdS-Schwarzschild. Any such black brane should be modular S-related to the singular solution. But that means the ``black brane'' will be horizonless and singular, and if e.g. $\alpha'$ effects resolve the singularity and pop out a horizon, then the entropy should be proportional to some power of $\alpha'$. But the ground state energy is a boundary term and is not proportional to $\alpha'$. This is inconsistent, by the Cardy formula which relates the two.

We consider our theory at inverse temperature $\beta$ on a spatial torus of side lengths $L_i$. The Euclidean solutions with the correct periodicity conditions are the toroidally compactified Poincar\'e patch, black brane, and $d-1$ AdS solitons
\begin{align}
 ds^2_{\textrm{pp}}&=r^2dx_0^2+\f{dr^2}{r^2}+r^2 d\phi_i d\phi^i\,,\\
ds^2_{\textrm{bb}} &= r^2\left(1-(r_h/r)^{d}\right)dx_0^2+\f{dr^2}{r^2\left(1-(r_h/r)^{d}\right)}+r^2d\phi_i d\phi^i\,,\\
ds^2_{\textrm{sol,k}} &= r^2dx_0^2+\f{dr^2}{r^2\left(1-(r_{0,k}/r)^{d}\right)}+r^2\left(1-(r_{0,k}/r)^{d}\right) d\phi_k^2+r^2 d\phi_j d\phi^j\,,
\end{align} 
all of which have the identification $x_0 \sim x_0 + \beta$. There are $d-1$ AdS solitons since there are $d-1$ circles that are allowed to pinch off in the interior. This means that we are picking supersymmetry-breaking boundary conditions around all cycles, which is motivated by maintaining $S$-invariance of our thermal partition function. 

The parameter $r_h$ ($r_{0,k}$) is fixed by demanding the $x_0$ ($\phi_k$) circle caps off smoothly:
\be
r_h= \frac{4\pi}{d \beta}\,,\qquad r_{0,k} =  \frac{4\pi}{d L_k}\,.
\ee
Considering the ensemble at finite temperature and zero angular velocity, we need to compare the free energy of these solutions:
\be
F_{\textrm{bb}}= -\frac{ r_h^d V_{d-1} }{16\pi G }\,, \qquad   F_{\textrm{sol,k}}= -\frac{r_{0,k}^d V_{d-1} }{16\pi G } \,,\qquad F_{\textrm{pp}} = 0\,.
\ee
The Poincar\'e patch solution never dominates so we will not consider it in what follows. We will also assume that the AdS soliton of minimal energy gives the vacuum energy of the theory under a toroidal compactification \cite{Horowitz:1998aa}.
\subsection{Thermal phase structure}
We will first consider the thermal phase structure, which can be illustrated by fixing a spatial torus and varying the inverse temperature $\beta$. The AdS soliton with the cycle of smallest length $L_{\text{min}}$ pinching off has minimal free energy and dominates all the other ones. We will denote this as the $k=$ min soliton. Thus, the two relevant solutions are this $k=$ min soliton and the black brane. These two exhibit a thermal phase transition at $\beta=L_{\textrm{min}}$ with the black brane dominating the ensemble at high temperature $\beta < L_{\textrm{min}}$. The energy at the phase transition is 
\be
E \big |_{r_h=\frac{4\pi}{d L_{\textrm{min}}}}=-\partial_\beta  \ \mathrm{log}  Z  =- (d-1) E_{\va}\,,
\ee
where $E_{\text{sol},k=\text{min}} = F_{\textrm{sol},k=\text{min}}=E_\va$ is the vacuum energy of the theory.

\subsection{Quantum phase structure}
A very important new feature in the phase structure of higher-dimensional toroidally compactified AdS spacetime is the existence of quantum phase transitions. These are phase transitions that can occur at zero temperature and are therefore driven by quantum fluctuations and not thermal fluctuations. They occur as we vary the spatial cycle sizes and reach a point where two spatial cycle sizes coincide and are minimal with respect to the rest. Let us call these cycle lengths $L_1$ and $L_2$ and pass from $L_1<L_2$ to $L_1>L_2$. In this case the vacuum energy exhibits a sharp transition from the $k=1$ soliton to the $k=2$ soliton. This is precisely the behavior that fixes $\tf=0$, as alluded to earlier. To exhibit a phase transition in the free energy instead of the vacuum energy, we need to restrict ourselves to the low-temperature phase $\beta>L_{\text{min}}$ where the black brane does not dominate.

\section{Necessary and sufficient conditions for universality}\label{proofs}
In this section we would like to highlight a few difficulties in generalizing a discussion from two dimensions to higher dimensions. Let us first consider a two-dimensional CFT with cycle lengths $\beta$ and $L$. For such a theory, vacuum domination of the torus partition function in channel $L$, for arbitrary cycle size $L>\beta$, is necessary and sufficient for universality of the partition function for all $\beta$. To see this, we write vacuum domination in the $L$ channel as
\be
Z(\beta)_L=Z(L)_\beta = \sum_{E_\beta} \exp\left(-L E_\beta\right)\approx \exp\left(-LE_{\va,\beta}\right) = \exp\left(\f{\pi  c L}{6\beta}\right)\,.
\ee
Due to the fact that the vacuum energy for two-dimensional CFT is uniquely fixed by conformal invariance, we get a universal answer for the partition function. In the $\beta$ channel, this form is that of an extensive free energy, and gives the Cardy formula in the canonical ensemble $S(\beta) = \pi c L/(3\beta)$. 

In higher dimensions, vacuum domination of the torus partition function in one channel seems neither necessary nor sufficient for extensive Cardy growth in a different channel. This is because the vacuum energy on a generic torus is not uniquely fixed by conformal invariance. But it turns out we can use $SL(d,\mathbb{Z})$ invariance to show that a slightly modified version of the statement is valid. In particular, we will show that vacuum domination in all channels except that of the smallest cycle is necessary and sufficient for universality of the partition function for all $\beta$. Before we begin, we will prove some useful properties of the function $f(\y)$ which characterizes the subextensive corrections to the vacuum energy and will play a starring role in our general CFT and symmetric orbifold analyses. Sections \ref{modularconstraints} and \ref{necsuff} will contain results about generic modular-invariant CFTs. Sections \ref{fneq0} and \ref{feq0} will then specify to large-$N$ theories. 

\subsection{Modular constraints on vacuum energy \label{modularconstraints}}
We now utilize the connection between the vacuum energy and the excited states implied by modular invariance, as first pointed out in appendix A of \cite{Shaghoulian:2015kta}. We will find that, somewhat surprisingly, modular invariance constrains all subextensive corrections to the vacuum energy to have a fixed sign and monotonic behavior. 

Consider a spatial torus with side lengths $L_1\leq \dots\leq L_{d-1}$ and take the quantization along $\beta$ at low temperature, which efficiently projects to the vacuum:
\be
\lim_{\beta\rightarrow \infty}\, \f{\log Z(\beta)_{\mathcal{M}_0}}{\beta} = - E_{\va,\mathcal{M}_0}=\f{\evac V_{\mathcal{M}_0}}{L_1^d}(1+f(\y))\,.
\ee
We also consider the $d-2$ quantizations $L_2,\dots,L_{d-1}$, which give
\be
\lim_{\beta\rightarrow \infty}\,\f{\log Z(L_i)_{\mathcal{M}_i}}{\beta} = \f{\evac V_{\mathcal{M}_0}}{L_1^d}(1+f(\y \setminus y_i, 0))+\lim_{\beta\rightarrow \infty}\,\f{1}{\beta}\log\left(\sum_{E} e^{-L_i(E-E_\va)_{\mathcal{M}_i}}\right),
\ee
where $\y \setminus y_i$ is the vector $\y$ without the $y_i$-th element. The reason for the different arguments of $f$ is that in the $L_i$ quantization,  instead of the ratio $L_1/L_i$ we have $L_1/\beta =0$ as $\beta \rightarrow \infty$. The second term on the right-hand-side does not vanish since the logarithm of the shifted partition function becomes linear in $\beta$ at large $\beta$ due to extensivity.

We want to analyze the monotonicity properties of $f(\y)$ with respect to its $d-2$ arguments. To analyze any given ratio $y_i$, we can equate the quantization along $\beta$ with the quantization along $L_i$. This gives
\be\label{keyvacs}
\f{\evac V_{\mathcal{M}_0}}{L_1^d}\left(f(\y)-f(\y\setminus y_i, 0)\right)=\lim_{\beta\rightarrow \infty}\,\f{1}{\beta}\log\left(\sum_{E} \exp\left(-L_i(E-E_\va)_{\mathcal{M}_i}\right)\right).
\ee
By unitarity, the right-hand-side is manifestly non-negative, so we conclude
\be
f(\y)-f(\y\setminus y_i, 0)\geq 0\,.
\ee
 Furthermore, the right-hand-side of \eqref{keyvacs} is a monotonically decreasing function of $L_i$. This means we can differentiate the left-hand-side with respect to $L_i$ and obtain
\be
f(\y)-f(\y\setminus y_i, 0)+L_i \p_{L_i} f(\y) \leq 0\implies \p_{L_i} f(\y) \leq 0\implies \p_{y_i}f(\y)\geq 0\,,
\ee
where the first implication follows from the previous positivity property. The second implication follows from the fact that increasing $L_i$ is the same as keeping all ratios $y_j$ fixed except for the ratio $y_i=L_1/L_i$, which is decreased. In particular, this means that the function increases under any possible variation. Furthermore, since $f(\mathbf{0})=0$ this means that $f(\y)\geq 0$. These facts will be used heavily in what follows.

Modular invariance can also be used to constrain the behavior of the vacuum energy under spatial twists. By re-interpreting the spatial twist as an angular potential in a different channel, we can see that the vacuum energy cannot increase due to a spatial twist. The proof goes as follows. Consider the following partition function in the low-temperature limit with twist $\theta_{kj}$ between two spatial directions $k$ and $j$:
\be
\lim_{\beta\rightarrow \infty}\f{\log Z(\beta ; \theta_{kj})_{\mathcal{M}_0}}{\beta} = - \EV({L_1, ..., L_{d-1}; \theta_{kj}}) \,.
\ee
Since the spatial directions are twisted, we may quantize along direction $k$, in which case the twist becomes an angular potential:
\bea
\lim_{\beta\rightarrow \infty}\f{\log Z(L_k; \theta_{kj})_{\mathcal{M}_k}}{\beta} =\lim_{\beta\rightarrow \infty}\f{1}{\beta}\log\left( \sum_E \exp\left( -L_k E_{\mathcal{M}_k} + i P_j \theta_{kj} \right)\right) \,.
\eea
The introduction of $\theta_{kj}$ only adds phases to the partition function in this channel, which decreases its real part. The vacuum energy is always manifestly real, so when equating the two quantizations it will be the case that the partition function with angular potential will evaluate to a real number. This means that the vacuum energy, which is negative, will be strictly greater or equal to its value without twists. This will be used in section \ref{section:orbifolds}.

\subsection{Necessary and sufficient conditions}\label{necsuff}
With the properties of the vacuum energy in hand, we are now ready to show that vacuum domination in all but the smallest channel is necessary and sufficient to have a universal free energy.

First we show \emph{sufficiency}. We consider an ordering $\beta<L_1\leq \dots \leq L_{d-1}$. Vacuum domination in the channels $L_i$ means
\be
Z(L_i)_{\mathcal{M}_i} = \exp\left(-L_i E_{\va, \mathcal{M}_i}\right)\,.
\ee
As we saw in the previous section, the vacuum energy is not uniquely fixed for higher-dimensional CFTs. However, equating the $d-2$ quantizations lets us extract the vacuum energy:
\begin{align}
Z(L_1)_{\mathcal{M}_1}&=Z(L_2)_{\mathcal{M}_2}=\dots = Z(L_{d-1})_\Md \\
\implies  -L_1 E_{\va, \mathcal{M}_1}&= -L_2 E_{\va, \mathcal{M}_2}=\dots = -L_{d-1} E_{\va, \mathcal{M}_{d-1}}\,.
\end{align}
Since $E_{\va, \mathcal{M}_i}$ is independent of $L_i$, we conclude that $E_{\va, \mathcal{M}_i}$ is linear in the cycle lengths $L_{j\neq i}$. The $\beta$ dependence is then fixed by dimensional analysis, and the coefficient is fixed by matching onto the asymptotic case of small $\beta$:
\be
E_{\va, \mathcal{M}_i}=-\evac V_{\mathcal{M}_i}/\beta^{d}.
\ee 
Thus, we see that vacuum domination in all but the smallest channel determines the functional form of the vacuum energy. We can now use $Z(\beta)_{\mathcal{M}_0} =Z(L_i)_{\mathcal{M}_i} $ to get 
\be
Z(\beta)_{\mathcal{M}_0} = \exp\left(\evac  L_iV_{\mathcal{M}_i}/\beta^{d}\right)=\exp\left(\evac  V_{\mathcal{M}_0}/\beta^{d-1}\right)\,.
\ee
This is just the Cardy formula. In a regular CFT it holds only asymptotically in small $\beta$, but here we have shown that vacuum domination in the spatial channels $L_i$ is sufficient to make it valid for all temperatures $\beta < L_i$.  
For $\beta>L_1$ we again have a universal expression for $Z(\beta)_{\Mz}$, which by assumption is given by the contribution of the vacuum only.

Showing that vacuum domination in all but the smallest cycle is \emph{necessary} for universality requires the properties of $f(\y)$ proven in the previous subsection. Consider the quantization along an arbitrary channel of cycle size $L_i$
\be
Z(L_i)_\Mi = \sum e^{-L_i E_\Mi} = e^{-L_i E_{\va, \Mi}} \sum e^{-L_i(E-E_0)_\Mi} \,.
\ee
In two spacetime dimensions, it is the vacuum contribution in this channel that gives Cardy behavior in the $\beta$ channel and therefore universality. The excited states contribute as positive numbers, and would ruin the Cardy behavior. Therefore it is necessary that they not contribute, i.e. necessary that we are vacuum dominated in this channel. In higher dimensions, one may worry that the excited state contributions cancel against the non-universal pieces of the vacuum energy, precluding the necessity of vacuum domination. However, by the positivity  of $f(\y)$ this can never happen. Thus, to get the correct Cardy behavior in the $\beta$ channel it is necessary that the excited states do not contribute. This is true for arbitrary channel $i$. We conclude that it is necessary to be vacuum dominated in all but the smallest cycle.

It is interesting that for a universal free energy it is necessary and sufficient to have vacuum domination in all but the smallest channel. One could have suspected that explicit assumptions about the subextensive corrections to the vacuum energy would have to enter, but they do not. 

We can state an equivalent set of necessary and sufficient conditions. To obtain a universal free energy for all $\beta$ on an arbitrary rectangular torus, it is necessary and sufficient to have vacuum domination in the largest spatial cycle, with the vacuum energy taking the universal form with no subleading corrections. In fact, by using the non-negativity and monotonicity of the subextensive corrections, we can state the necessary and sufficient condition as vacuum domination in the largest spatial cycle, with the vacuum energy on a square torus of side length $L$ equal to $\evac/L$.

In the rest of this section we will restrict attention to large-$N$ theories. 

\subsection{Sparseness constraints without assuming $\tilde{f}(\y)=0$}\label{fneq0}
It is difficult to make progress in the case where we make no explicit assumptions about the functional form of the vacuum energy. To achieve vacuum domination in all but the smallest channel of a large-$N$ theory, we can bound the entire spectrum on an arbitrary spatial torus of side lengths $L_1\leq L_2\leq \dots \leq L_{d-1}$ as 
\be\label{cbound}
\rho(\D_\Mz)\lesssim \exp\left(L_1\D_\Mz\right)\,,\qquad \D_\Mz \equiv (E-\EV)_\Mz\,.
\ee
This is a necessary and sufficient condition, although it is possible that it is implied by a more minimal set of necessary and sufficient conditions. To see how this condition arises, one writes the partition function as
\be
Z(\beta)_\Mz = \exp\left(-\beta \EV\right) \sum \exp\left(-\beta\D_\Mz\right)\rho(\D_\Mz)
\ee
and bounds the density of states as \eqref{cbound} for the entire spectrum. At large $N$, with a vacuum contribution that scales exponentially in $N$, this suppresses all excited state contributions as soon as $\beta > L_1$. This means all cycles except the smallest will be vacuum-dominated, as required. We give another method of proof for vacuum domination in appendix \ref{alternate} which restricts the sparseness bound to only the light states, but requires an additional assumption on the field theory.

We can also show that it is necessary and sufficient to solve the problem on a spatial square torus, i.e. that the free energy is universal for all $\beta$ on a spatial square torus of side length $L$. The necessary direction is obvious. To show sufficiency, consider the quantization along $L$:
\begin{align}
Z(L)_{\Md} &=\exp\left(-LE_{\va,\Md}\right) \sum_\D \exp\left(-L\D_{\Md}\right)\\
&= Z(\beta)_{\Mz} \approx \exp\left(\evac L^{d-1}/\beta^{d-1}\right)\,.
\end{align}
where the final expression is by assumption of universality. The only way to satisfy this equality is for the contribution of the excited states and the subextensive corrections to the vacuum energy in the $L$ channel to vanish. In particular we are vacuum dominated in the $L$ channel. Taking arbitrary $L_{d-1}>L$ keeps us vacuum dominated since it is at even lower temperature:
\begin{align}
Z(L_{d-1})_\Md \approx \exp\left(\evac L_{d-1} L^{d-2}/\beta^{d-1}\right)\,.
\end{align}
In the $\beta$ channel this gives us the ordinary Cardy formula with no subextensive corrections, and in another $L$ channel we have
\begin{align}
Z(L)_{\mathcal{M}_{d-2}} &= \exp\left(-LE_{\va, \mathcal{M}_{d-2}}\right) \sum_\D \exp\left(-L\D_{\mathcal{M}_{d-2}}\right)\\
&= Z(L_{d-1})_\Md \approx \exp\left(\evac L_{d-1} L^{d-2}/\beta^{d-1}\right)\,.
\end{align}
Again, this means that we are vacuum dominated in the $L$ channel. Now we can consider arbitrary $L_{d-2}$ satisfying $L<L_{d-2}<L_{d-1}$, for which we will remain vacuum dominated:
\be
Z(L_{d-2})_{\mathcal{M}_{d-2}}\approx \exp\left(\evac L_{d-1}L_{d-2} L^{d-3}/\beta^{d-1}\right)\,.
\ee
By equating this expression with the partition function in the $L_{d-1}$ channel, we see that we are still vacuum dominated in that channel. By continuing this procedure we are able to generalize to an arbitrary torus $\beta < L_1 < \dots < L_{d-1}$, and we obtain 
\be
\log Z(\beta) = \left\{\begin{array}{cc}
\evac V_\Mz/\beta^{d-1},\qquad \beta < L_1\\
\evac V_{\mathcal{M}_1}/L_1^{d-1},   \qquad \b > L_1
\end{array}\right..
\ee
 Altogether, we have that the free energy is universal at all temperatures on an arbitrary spatial torus. So solving the problem on a spatial square torus is both necessary and sufficient to solving the general problem, thanks to properties of the positivity of $f(\y)$. 
\subsection{Sparseness constraints assuming $\tf=0$}\label{feq0}
In this section we will show that assuming $\tf=0$ (up to $1/N$ corrections) allows us to exhibit a constraint on the light spectrum that naturally generalizes the two-dimensional case. This is not too surprising, as $\tf=0$ is automatically true in two dimensions, although some more work will be required in higher dimensions.

We start by considering the special torus with ordering $\b < L < L^2/\b < \dots < L^{d-1}/\b^{d-2}$. As discussed in the introduction, this special torus has an exact low-temperature/high-temperature duality $Z(\beta)_{\mathcal{M}_0} = Z(L^d/\beta^{d-1})_{\mathcal{M}_0}$. This will allow us to uplift the arguments of \cite{Hartman:2014oaa} to our case. In the upcoming manipulations, we will not keep explicitly the specification of the spatial manifold $\mathcal{M}_0$, since this duality allows us to keep our spatial manifold fixed once and for all.

By following the steps in \cite{Hartman:2014oaa}, one can show that the partition function is dominated by the light states up to a theory-independent error. We will denote light states as those with energy $E<\epsilon$ for some arbitrary $\epsilon$. We have
\be
\log Z_{\textrm{light}}(L^d/\b^{d-1}) \leq \log Z(\b) \leq \log Z_{\textrm{light}}(L^d/\b^{d-1}) - \log\left( 1-e^{\e(\b - L^d/\b^{d-1})} \right).
\ee 
This error grows arbitrarily large as $\b \to L$ or $\epsilon\to 0$. For $\b > L$ we can derive a similar upper and lower bound. 

For a family of CFTs labeled by $N$, we assume that the vacuum energy also scales with $N$. This will be true in all examples we consider. When taking $N$ large, we can scale $\epsilon\rightarrow 0$, in which case the partition function is squeezed by its bounds and given just by the light states up to $\mathcal{O}(1)$ corrections. In the context of assuming $\tf=0$, we then obtain universality
\be\label{vacdomst}
\log Z(\b) = \left\{\begin{array}{cc}
\log Z_{\textrm{light}}(L^d/\b^{d-1}) = -\frac{L^d}{\b^{d-1}}E_\va & \b < L\\
\log Z_{\textrm{light}}(\b) = -\b E_\va & \b > L
\end{array}\right.,
\ee
if and only if the density of light states is bounded as 
\be\label{bound}
\rho(\D) \lesssim \exp \left( \frac{L^d}{\b^{d-1}}\D\right), \qquad \D \leq  - E_\va\,,
\ee
where $\D = E- E_\va$. Notice that if we did not assume a universal form for the vacuum energy with $\tf=0$, the free energy would still be very theory-dependent.

To generalize the argument above to an arbitrary $d$-torus, the idea will be to push the special torus very close to the square torus. From here, we can use the fact that whenever a partition function is dominated by the vacuum contribution at some inverse temperature $\b$, then it will also be dominated by that contribution for larger $\b$. Channel by channel, we will see that we will be able to generalize to an arbitrary torus. Assuming a universal form of the vacuum energy will be crucial for this argument.

It will be convenient to consider starting with a quantization along the $L^{d-1}/\b^{d-2}$ channel, because it is the largest cycle when $\b < L$. We will now restore the explicit spatial manifold dependence since we will be considering quantizations along different channels. We have
\be
Z(L^{d-1}/\b^{d-2})_\Md = Z(L^d/\b^{d-1})_\Mz = Z(\b)_\Mz \,.
\ee
By using \eqref{vacdomst} we can write this as 
\be
Z(L^{d-1}/\b^{d-2})_\Md =  \exp\left(-\f{L^d}{\b^{d-1}}E_{\va, \Mz}\right)= \exp\left(-\f{L^{d-1}}{\b^{d-2}}E_{\va, \Md} \right)\,.
\ee
This means that we are vacuum-dominated in the $L^{d-1}/\beta^{d-2}$ channel. 

Let us now take a larger cycle $L_{d-1} > L^{d-1}/\b^{d-2}$, for which we will remain vacuum-dominated:
\be
Z(L_{d-1})_\Md = \exp\left(-L_{d-1}E_{\va, \Md}\right)\,.
\ee
Quantizing now along the the second largest cycle $L^{d-2}/\b^{d-3} < L_{d-1}$ gives us
\be\label{Ld1quant}
Z(L^{d-2}/\b^{d-3}) = \exp\left(-L^{d-2}/\b^{d-3}E_{\va, \mathcal{M}_{d-2}}\right)\sum_{\D} \exp\left(-L^{d-2}\D_{\mathcal{M}_{d-2}}/\b^{d-3}\right).
\ee
But by our assumption $\tf=0$, we have
\be
L_{d-1}E_{\va, \Md} = L^{d-2} E_{\va, \mathcal{M}_{d-2}}/\beta^{d-3} \,,
\ee
which means that $Z(L^{d-2}/\b^{d-3})$ is given by its vacuum contribution only. One can now consider $L_{d-2}>L^{d-2}/\beta^{d-3}$, for which we will remain vacuum-dominated in the $L_{d-2}$ channel. By comparing to the $L_{d-1}$ channel, we can verify that we remain vacuum-dominated there as well. We can now move to the $L^{d-3}/\beta^{d-4}$ channel and continue this procedure up to and including the $L$ channel. In a final step, we can compare to the $\beta$ channel and see that it indeed has universal Cardy behavior:
\be
\log Z(\beta) = \f{\evac V_\Mz}{\beta^d}\,.
\ee
There is no need now to consider smaller $\beta$ since we have already considered general variations of the other $d-1$ cycles. Since the partition function is a function of $d-1$ independent dimensionless ratios, we have already captured all possible variations.

The generality of the torus that results from this procedure is restricted by the special torus with which we began. But notice that the special torus can be arbitrarily close to a $d$-dimensional square torus, which means this procedure results in a universal free energy on an arbitrary torus. From this argument it is clear that the only assumption made on the spectrum is the bound in \eqref{bound}. In fact, it is enough to impose this constraint for the square torus, since our procedure begins from that case (or arbitrarily close to it) and generalizes to an arbitrary torus. The sparseness constraint is therefore
\be\label{finalbound}
\rho(\D) \lesssim \exp\left( L\D_{L\times L \times \dots \times L} \right) 
\ee
and is imposed only on the states with energies $E=\D+E_\va <0$. 

\section{Symmetric Product Orbifolds in $d>2$}\label{section:orbifolds}
In this section we construct orbifold conformal field theories in higher dimensions using a procedure analogous to the one in two dimensions. We will see that these theories contain both twisted and untwisted sector states and will give an estimate for the density of states within these sectors. Finally, we will show that under the assumption that $\tf=0$, the free energy has a universal behavior at large $N$ which agrees with Einstein gravity.

\subsection{A review of permutation orbifolds in two dimensions}
In two dimensions, symmetric product orbifolds (or the more general permutation orbifolds) provide a vast landscape of two-dimensional CFTs with large central charge that have a potentially sparse spectrum and are thus of interest in the context of holography \cite{Belin:2015hwa, Haehl:2014yla, Belin:2014fna, Benjamin:2015vkc}. The goal of this section will be to extend these constructions to higher dimensions. We start by a review of permutation orbifolds in two dimensions which will set most of the notation that we will then carry over to higher dimensions. Permutation orbifolds are defined by the choice of two parameters: a ``seed" CFT $\mathcal{C}$ and a permutation group $G_N \subset S_N$. A permutation orbifold $\mathcal{C}_N$ is then defined to be
\be
\mathcal{C}_N \equiv \frac{\mathcal{C}^{\otimes N}}{G_N}\,.
\ee
The procedure by which we take this quotient is called an orbifold. It projects out all states of the product theory that are not invariant under the action of the group. The Hilbert space thus gets restricted to
\be
\mathcal{H}^{\otimes N} \  \longrightarrow \ \frac{\mathcal{H}^{\otimes N}}{G_N}\,,
\ee
where $\mathcal{H}$ is the Hilbert space of $\mathcal{C}$. This projection onto invariant states is crucial as it gets rid of most of the low-lying states and hence provides some hope of obtaining a sparse spectrum. When computing the torus partition function, this projection onto invariant states is implemented by a sum over all possible insertions of group elements in the Euclidean time direction. This is summarized by the following formula
\be
Z_{\text{untw}}=\frac{1}{|G_N|}\sum_{g\in G_N}\begin{array}{r}\\  g~  \begin{array}{|c|}\hline ~~~ \\ \hline \end{array} \\  \phantom{g} \end{array} \label{untwZ}
\ee
where the box represents the torus with the vertical direction being Euclidean time.

However, \rref{untwZ} is obviously not modular invariant as it singles out the time direction. Modular invariance is restored in the following way
\be
Z_{\text{tot}}=\frac{1}{|G_N|}\sum_{g,h \in G_N | gh=hg}\begin{array}{r}\\  g~  \begin{array}{|c|}\hline ~~~ \\ \hline \end{array} \\  h~ \end{array} \label{untwZ} \,.
\ee
The requirement that the two group elements must commute comes from demanding that the fields have well-defined boundary conditions \cite{Ginsparg:1988ui}. The insertion of elements $h$ in the spatial direction are interpreted as twisted sectors, where the boundary conditions of the fields are twisted by group elements. There is one twisted sector per conjugacy class of the group, which in the case of $G_N=S_N$ gives one twisted sector per Young diagram. In \cite{Belin:2015hwa, Haehl:2014yla, Belin:2014fna}, the space of permutation orbifolds was explored and a criterion was given for these theories to have a well-defined large $N$ limit (and thus a potential holographic dual). It was found that many properties of the spectrum depends solely on the group $G_N$ and not on the choice of the seed theory. Groups that give a good large-$N$ limit are called oligomorphic permutation groups \cite{MR1066691,MR2581750,MR0401885}. Although a complete proof is still missing, it is believed permutation orbifolds by oligomorphic groups all have at least a Hagedorn density of light states, but the growth may be even faster \cite{Belin:2015hwa, Belin:2014fna}. For the symmetric group, it was shown in \cite{Keller:2011xi, Hartman:2014oaa} that the growth is exactly Hagedorn with the precise coefficient saturating the bound on the density of light states produced in \cite{Hartman:2014oaa}.

Symmetric product orbifolds thus reproduce the phase structure of 3d gravity. Note that they are still far from local theories of gravity such as supergravity on $AdS_3 \times S^3 $, as their low-lying spectrum is Hagedorn and so they look more like classical string theories. The D1-D5 CFT has a moduli space that is proposed to contain a point, known as the orbifold point, where the theory becomes a free symmetric product orbifold theory. According to this proposal, the orbifold point is connected to the point where the supergravity description is valid by an exactly marginal deformation. It is only the strongly coupled theory that is dual to supergravity, and from this point of view it is surprising that the free theory realizes the phase structure of gravity.

\subsection{Symmetric product orbifolds in higher dimensions}
In two dimensions, we saw that symmetric product orbifolds are examples of theories with a sparse enough spectrum to satisfy the bound from \cite{Hartman:2014oaa} and thus have a universal phase structure at large $N$. We would now like to construct weakly coupled examples of theories satisfying our new criteria in higher dimensions. In dimensions greater than two, it is in general much harder to construct large-$N$ CFTs. One may of course take tensor products but these will never have a sparse enough spectrum. In fact, the spectrum below some fixed energy level will not even converge as $N\to\infty$. Imposing some form of Gauss' law to project out many of the low-lying states is usually done by introducing some coupling to a gauge field, which makes preserving conformal invariance highly non-trivial. A natural way to achieve this same projection is through the construction of orbifold conformal field theories familiar from two dimensions. To the best of our knowledge, there is no construction of orbifold conformal field theories in higher dimensions, which as explained in the previous subsection is perhaps the most natural way of obtaining theories that are conformal, have a large number of degrees of freedom, but also a sparse low-lying spectrum. 

We will now describe the construction of symmetric product orbifolds in $d$ dimensions.\footnote{Here we will assume that the group is $S_N$ but the generalization to other permutation groups follows trivially from our construction.} We will construct the partition function, i.e. the Hilbert space and the spectrum of the Hamiltonian on $\mathbb{T}^{d-1}$. We comment on other properties of the theory such as correlation functions in the discussion section. 

The starting point is again to consider a seed CFT$_d$ $\mathcal{C}$ and to define the orbifold theory $\mathcal{C}_N$ as
\be
\mathcal{C}_N \equiv \frac{\mathcal{C}^{\otimes N}}{S_N}
\ee
The orbifolding procedure goes as follows. We start by projecting onto invariant states by inserting all elements of the group in the time direction. This gives
\be
Z_{\text{untw}} = \frac{1}{N!}\sum_{g \in S_N } \cube{g}{\phantom{h}}{\phantom{k}} \,.
\ee
The box of the 2d case has now been lifted to a $d$-dimensional hypercube which again describes the torus. We will represent it by a $3d$ cube and leave the other dimensions implicit. Again, the mere projection is obviously not modular invariant. By applying elements of $SL(d,\mathbb{Z})$ (for instance the $S$ element given in \rref{sldgen}), we quickly see that group elements must also be inserted in the space directions. Having well-defined boundary conditions for the fields constrains the $d$ group elements to be commuting. The partition function of the orbifold theory is then defined as
\be
Z_{\text{orb}} = \frac{1}{N!}\sum_{\substack{g_0,...,g_{d-1} \in S_N \\ g_ig_j=g_jg_i \forall i,j}} \cube{g_0}{g_1}{\ \ \ g_{d-1}} \label{orbifoldZ} \,.
\ee
Twisted sector states will correspond to any states with non-trivial insertions in any of the space directions. The different twisted sectors are no longer labeled just by conjugacy classes, but by sets of $d-1$ commuting elements, up to overall conjugation. This orbifolding procedure describes a well-defined $SL(d,\mathbb{Z})$-invariant partition function. 


\subsection{Spectrum of the theory}
\subsubsection{The untwisted sector states}\label{untwisted}
We now turn our attention to the spectrum of these orbifold theories. 
Other properties will depend strongly on the choice of seed. We start by considering the untwisted sector states. These are given by states of the product theory, up to symmetrization. From the point of view of the partition function, their contribution consists of all elements in the sum \rref{orbifoldZ} where $g_1=...=g_{d-1}=1$. 
Consider the contribution of a $K$-tuple to the density of states. A $K$-tuple is a state where $K$ of the $N$ CFTs are excited, while the other $N-K$ are in the vacuum. The contributions of all possible $K$-tuples of distinct states are encapsulated by the following expression:
\be
\rho(\Delta) = \int dK \int d\Delta_1....d\Delta_K \f{1}{K!}\rho_0(\Delta_1).... \rho_0(\Delta_K) \delta ( \Delta- \sum_{j=1}^K \Delta_j)\,,
\ee
where $\Delta=E-N\EV$, $\Delta_i=E_i-\EV$ and $\rho_0$ is the density of states of the seed theory.\footnote{Here we use the notation that $\Delta$ is a shifted energy that satisfies $\D\geq 0$, but we wish to emphasize that it is \emph{not} in any way related to the scaling dimension of a local operator.} It can be shown that the contribution of $K$-tuples with subsets of identical states do not give a larger contribution than the one considered here, so it is sufficient to focus on this case. The combinatorial prefactor $1/K!$ was introduced to remove the equivalent permutations of the $K$ states. One way to understand its inclusion is to consider how the orbifold projection is done. A given $K$-tuple in the product theory is made $S_N$ invariant by summing over all of its possible permutations. For example, the $3$-tuples $\{a,b,c\}, \{a,c,b\}, \{b,a,c\},\{b,c,a\}, \{c,a,b\}, \{c,b,a\}$ of the pre-orbifolded theory lead to the same orbifolded $3$-tuple and thus should only be counted once. The triple integral giving $\rho(\Delta)$, when left to its own devices without combinatorial prefactor, would count all six configurations.

Along with the states being distinct, let us first assume that each of the individual degeneracies can be approximated by the Cardy formula of the seed theory. The Cardy formula in higher dimensions was given in \rref{cardyhigherd} and reads
\be
\log \rho(E) = \frac{d}{(d-1)^{\frac{d-1}{d}}} (\evac V_{d-1})^{\frac{1}{d}} E^{\frac{d-1}{d}}\,. \label{cardyformula}
\ee
Now let us proceed as in \cite{Belin:2015hwa} to find the density of states. Performing the integrals over energies $E_i$ by a saddle-point approximation where the large parameter is the total energy $E$, we find saddle-point values $E_i=E/K$ for all $i$. To assure that the state in each copy is distinct, we need the degeneracy to pick from to be much larger than $K$. Thus the validity of this assumption and the validity of the Cardy formula in each seed theory require, respectively,
\begin{equation}
\exp \left[ \frac{d}{(d-1)^{\frac{d-1}{d}}} (\evac V_{d-1})^{\frac{1}{d}} (\Delta/K+E_{\text{vac}})^{\frac{d-1}{d}} \right] \gg K, \quad \Delta/K \gg |E_\text{vac}|\,.
\end{equation}
We will check whether these conditions are satisfied at the end. Note that the second constraint implies that we can drop $E_{\text{vac}}$ in the Cardy formula when expressed in terms of $\Delta$. We thus have
\be
\rho(\Delta) \sim \int dK \exp \left[{ da K^{\frac{1}{d}}  \Delta^{\frac{d-1}{d}}-K \log K + K}\right]
\ee \label{adef}
with
\be
a\equiv\frac{1}{(d-1)^{\frac{d-1}{d}}} (\evac V_{d-1})^{\frac{1}{d}}\,. \label{adef}
\ee
We can now do a second saddle-point approximation to evaluate the integral over $K$. The large parameter is again given by the total shifted energy $\Delta$. The saddle point equation is
\be
a\Delta^{\frac{d-1}{d}} K_s^{\frac{1-d}{d}}-\log K_s =0\,,
\ee
which gives
\be
K_s\sim \frac{a^{\frac{d}{d-1}}\Delta}{\left(\log \left[a^{\frac{d}{d-1}}\Delta\right]\right)^{\frac{d}{d-1}}}
\ee
at large $\D$. Plugging this back in the density of states we find
\be
\rho(E) \sim \exp \left[ (d-1)  \frac{a^{\frac{d}{d-1}}\Delta}{ \left(\log \left[a^{\frac{d}{d-1}}\Delta\right]\right)^{\frac{1}{d-1}}} \right],
\ee
where we have used large $\Delta$ to drop subleading pieces which either have a larger power of the logarithm in the denominator or are terms proportional to $\log \log \Delta$. We find a growth of states that is slightly sub-Hagedorn and the growth increases with the dimension of the field theory.
Inserting $K_s$ in our necessary assumptions shows that they can be satisfied for large enough $\Delta$. In particular, the second condition becomes 
\be
a^{\frac{d}{d-1}}\Delta\gg \exp\left[ a \, |E_\text{vac}|^{\frac{d-1}{d}}\right]
\ee
which is then sufficient to satisfy the first condition. Here $\EV$ is the vacuum energy of the seed theory and does not scale with $N$. Notice also that $K_s$ grows with $\D$ and must not violate the bound $K_s\leq N$. This implies a bound on our energies from the saddle:
\be
a^{\frac{d}{d-1}}\Delta \lesssim N\left[\log(N)\right]^{\frac{d}{d-1}}\,.
\ee
So altogether our density of states formula is reliable in the range
\be
\exp\left[ a \, |E_\text{vac}|^{\frac{d-1}{d}}\right] \ll a^{\frac{d}{d-1}}\Delta  \lesssim N\left[\log(N)\right]^{\frac{d}{d-1}}\,.
\ee
In particular we can consider energies that scale with $N$. However, as we will shortly see, the density of states quickly becomes dominated by the twisted sectors. Note that this growth of states is also a lower bound for any permutation orbifold as orbifolding by a subgroup of $S_N$ always projects out fewer states.

\subsubsection{The twisted sector states}
We will now give a lower bound on the density of states coming from the twisted sectors. If the intuition from two dimensions carries over, it will be the twisted sectors that give the dominant contribution to the density of states. Indeed, this is the result we will find. We start by a more general discussion of twisted sector states and their contribution to the partition function.

A twisted sector is given by $d-1$ commuting elements $g_1,...,g_{d-1}$ of $S_N$, up to overall conjugation. There is also a projection onto $S_N$-invariant states by summing over elements in the time direction but at this point we only focus on the identity contribution in that direction. We define $\mathbb{T}$ to be the original $d$-torus used to compute the partition function. We leave the dependence on the vectors $U_0,...,U_{d-1}$ implicit. Let us consider the action of the subgroup $G_{g_1,...,g_{d-1}}$ of $S_N$ (defined to be the group generated by $g_1,...,g_{d-1}$) on the $N$ copies of the CFT. The action of this group will be to glue certain copies of the CFT together. Concretely, let $\Phi^k$ denote a field on $\mathbb{T}$ of the $k$-th CFT, then in the twisted sector defined by $G_{g_1,...g_{d-1}}$ this field has boundary conditions
\be
\Phi^{k}(x_0,x_1, ... , x_j + L_j, ... , x_{d-1}) = \Phi^{g_j(k)}(x_0,x_1 , ..., x_j, ... , x_{d-1})\,.
\ee
Tracking the orbit of the $k$-th copy under $G_{g_1,...g_{d-1}}$ allows us to define a single field $\widetilde{\Phi}$ with modified boundary conditions. In particular it will have larger periods. A field $\widetilde{\Phi}^i$ can be defined for each orbit of the group $G_{g_1,...g_{d-1}}$ and we will denote the set of these orbits by
\be
\{ O_i\}\,, \ \ \ i=1,... i_{\text{max}}\,,
\ee
where $i_{\text{max}}$ depends on the precise choice of $g_1,...,g_{d-1}$. As the different orbits do not talk to each other, the path integral will split into a product of $ i_{\text{max}}$ independent path integrals, one over each field $\widetilde{\Phi}^i$. The new boundary conditions of the fields in a given $O_i$ under the action of $G_{g_1,...,g_{d-1}}$ enable us to rewrite that particular contribution to the path integral as a torus partition function, but now with $\mathbb{T}$ replaced by a new torus $\widetilde{\mathbb{T}}_i$. The original identifications coming from \rref{Umatrix} were
\be
(x_0,x_1,..,x_{d-1}) \sim (x_0,x_1,..,x_{d-1}) + \sum_{i=0}^{d-1} n_i U_i \,.
\ee
for any integers $n_i$. Once the elements $g_1,...,g_{d-1}$ are inserted the identifications are changed and they are encoded in a new torus. As these boundary conditions follow from the orbits, the identifications from the new torus are given by the elements in $G_{g_1,...,g_{d-1}}$ that leave the orbit invariant, i.e.
\be
g_1^{m_1}...g_{d-1}^{m_{d-1}} O_i=O_i \,.\label{cyclecondition}
\ee
This means that the identifications become
\be
(x_0,x_1,..,x_{d-1}) \sim (x_0,x_1,..,x_{d-1}) + \sum_{i=0}^{d-1} m_i U_i \,.
\ee
with the $m_i$ such that \rref{cyclecondition} is satisfied. Alternatively, one can define new vectors in the following way
\bea
\widetilde{U}_1&=&m_{1}^{\text{min}} U_1+m_{1,2} U_2 + ... + m_{1,d-1}U_{d-1}\,,\notag \\
&\vdots& \notag \\
\widetilde{U}_{d-2}&=&  m_{d-2}^{\text{min}} U_{d-2} + m_{d-2,d-1}U_{d-1}\,,\\
\widetilde{U}_{d-1}&=&  m_{d-1}^{\text{min}}U_{d-1} \,,\notag
\eea
where $m_{d-1}^{\text{min}}$ is the smallest integer $m_{d-1}$ such that $g_{d-1}^{m_{d-1}} O_i=O_i $, $(m_{d-2,d-1},m_{d-2}^{\text{min}})$ are the pair with smallest non-zero $m_{d-2}$ such that $g_{d-1}^{m_{d-2,d-1}} g_{d-2}^{m_{d-2}^{\text{min}}}O_i=O_i $ and the $(m_1^{\text{min}}, ...,m_{1,d-1})$ are the set of integers with minimal non-zero $m_1$ such that \rref{cyclecondition} is satisfied. These vectors define a new torus $\widetilde{\mathbb{T}}_i$ with volume
\be
\text{Vol}(\widetilde{\mathbb{T}}_i)= \left(\prod_j m_{j}^{\text{min}} \right) \text{Vol}(\mathbb{T}) \equiv |O_i| \text{Vol}(\mathbb{T})\,.
\ee
Since the $g_i$ commute, $|O_i|$ is just the number of elements in the orbit $O_i$.

A twisted sector will thus give a set of new tori $\widetilde{\mathbb{T}}_i$ whose different volumes depend on the orbits of the action of $G_{g_1,...,g_{d-1}}$. For each orbit of that action, we will get a separate torus and schematically, this will give a contribution to the partition function of the form
\be
Z_{\text{tot}}\sim\prod_{i} Z(\widetilde{\mathbb{T}}_i)\,,
\ee
where the product over $i$ is a product over the orbits. This is a generalization of Bantay's formula \cite{Bantay:1997ek} to higher dimensions. For every orbit $O_i$ we have
\be
\text{Vol}(\widetilde{\mathbb{T}}_i)= |O_i| \text{Vol}(\mathbb{T})\,,
\ee
where $|O_i|$ is the length of the orbit. We will now calculate the contribution to the partition function from a single non-trivial orbit of length $L =M^{d-1}$ giving a torus with equal rescaling $M$ in all spatial directions. For simplicity, we also consider a case with $m_{i,j}=0 \ \forall i\neq j$. The torus $\widetilde{\mathbb{T}}_i$ corresponding to this orbit is then
\be
(\widetilde{U}_0,...,\widetilde{U}_{d-1}) = (U_0, M U_1, ... , M U_{d-1} ) \,. \label{Tbigger}
\ee
We can always find elements $g_1,...,g_{d-1}$ that produce the desired torus with equal scaling of the spatial cycles. To produce the new torus given in \rref{Tbigger}, we use for example the following elements:
\bea
g_1&=&\left(1\ ...\ M\right)\left(M+1\ ... \ 2M\right) ... \ ( M^{d-1}-M+1 \ ...\ M^{d-1})(M^{d-1}+1)\ ... \ (N) \notag \\
g_2&=&\left(1\ \  M+1\ ...\ M(M-1)+1\right) ... \  \notag \\
&\phantom{=}&(M^{d-1}-M(M-1) \ \ \  M^{d-1}-M(M-2) \ ... \ M^{d-1})(M^{d-1}+1)\ ... \ (N) \notag \\
&\vdots&  \\
g_{d-1}&=& (1 \ \ M^{d-2}+1 \ ... \ M^{d-2}(M-1)+1) ...(M^{d-2} \ \ 2M^{d-2} \ ... \ M^{d-1})\ \notag \\
&\phantom{=}&(M^{d-1}+1)\ ... \ (N) \notag
\eea
for $L=M^{d-1}$. For example in $d=3$ and for $L=9$, we get
\bea
g_1&=&(1 \ 2 \ 3)(4 \ 5 \ 6)(7 \ 8 \ 9)(10)...(N)\,, \notag \\
g_2&=& (1 \ 4 \ 7)(2 \ 5 \ 8)(3 \ 6 \ 9)(10)...(N)\,.
\eea
One can quickly check that all these elements commute and that they define an orbit of length $L$ as well as $N-L$ singlets. One can also check that $m_1^\text{min}=...=m_{d-1}^\text{min}=M$. We will call $Z_{\text{sq}}$ this particular contribution to the partition function, and it reads
\bea
Z_{\text{sq}}&=&Z(U_0,U_1, ... , U_{d-1}) ^ {N-L} Z(U_0, M U_1, ... ,M U_{d-1}) \notag \\
&=&Z(U_0,U_1, ... ,U_{d-1}) ^ {N-L} Z(U_0/L^{\frac{1}{d-1}}, U_1, ..., U_{d-1})\,,
\eea
where we uniformly rescaled the torus and used $L=M^{d-1}$. From this, we can infer the behaviour of the density of states:
\begin{align}
Z_{\text{sq}}=\sum_E \rho_{\text{sq}}(E)e^{-\beta E} = e^{-\beta E_\text{vac} (N-L)}\left(1+\dots\right)\sum_E \rho_0(E)e^{-\beta E/ L^{\frac{1}{d-1}}}\,.
\end{align}
We can ignore the excited states encapsulated in ``\dots" as they will only increase $\rho_{\text{sq}}(E)$, which will increase our final answer. In this section, we are only after a lower bound for the density of states so we can ignore such terms. Shifting $E$ to $ L^{\frac{1}{d-1}}(E-E_\text{vac}(N-L))$ gives us
\be 
\rho_{\text{sq}}(E) = \rho_0( L^{\frac{1}{d-1}}(E-E_\text{vac}(N-L)))\,. \label{rhoshifted}
\ee
This will be the key formula to derive the final result.

In the full partition function we sum over all $L\leq N$ and for large $L$, we are in a regime where we may use the Cardy formula of the seed theory given in (\ref{cardyhigherd}). To find the twisted sector that gives the maximal contribution at energy $E$, we evaluate the sum over $L$ using a saddle point approximation. The resulting saddle point equation for $L$ is solved by
\be 
L_s = \frac{(E_\text{vac}N-E)}{dE_\text{vac}}\,, \label{Lsaddle}
\ee 
which will be a good approximation provided $L_s\gg 1 $. We now plug this back in \rref{rhoshifted} and use the Cardy formula \rref{cardyhigherd} to obtain
\be
\rho(E)\sim \exp \left[ a \frac{(d-1)^{\frac{d-1}{d}}}{|E_\text{vac} | ^{1/d}} (E- N E_\text{vac})\right]. \label{hagedorngrowth}
\ee
Note that this is a Hagedorn growth as in two dimensions but the coefficient of the Hagedorn growth depends on the vacuum energy of the seed theory. This is somewhat a loss of universality compared to two dimensions and it will be very important in what follows to understand precisely the properties of the vacuum energy of the orbifold theory. This will be the task of the next subsection. The regime in which this expression is reliable is for $1\ll L_s \leq N$ which in terms of energies is
\be
1 \ll \frac{E-NE_\text{vac}}{|E_\text{vac}|}\leq dN\,.
\ee
Finally, it is important to emphasize that this is merely a lower bound on the density of states\footnote{In fact, the method used in this section only gives an estimate for the lower bound. We have only inserted one element - the identity - in the time direction and have not taken into account the projection to $S_N$ invariant states. Following the method we will use in section \ref{freeenergyorbi} one can show that this estimate is actually precise.}. We have only given the contribution from one type of twisted sectors and other sectors might dominate. We have also not taken into account the projection onto $S_N$ invariant states by inserting commuting elements of the group in the time direction. In two dimensions, one can show that the estimate coming from this particular twisted sector (called long strings in $2d$) actually gives the dominant contribution. We will discuss this further when analyzing the free energy but we first turn our attention to the vacuum energy.

\subsection{Vacuum energy of the orbifold theory}
We want to understand precisely the properties of the vacuum energy of the orbifold theory. In two dimensions, it is clear that the central charge gets multiplied by $N$ when going from the seed theory to the product (or orbifold) theory. Since the vacuum energy is fixed by the central charge, it also gets multiplied by $N$. Naively, one would expect a similar behavior in higher dimensions. The all-vacuum contribution in the untwisted sector indeed has energy $N \EV$, but it may be possible that other twisted sectors give even more negative contributions. We will now address this possibility and show that it is impossible, so that the vacuum energy of the orbifold theory is in fact given as 
\be
\EV^{\text{orbi}}=N \EV\,. \label{evacN} 
\ee
To prove this, first recall that it is not necessary to consider twisted sectors inducing twists between any of the dimensions because they always increase the vacuum energy, as explained in section \ref{modularconstraints}. The only thing we need to check is that rescalings of the torus do not give a contribution that is more negative than \rref{evacN}. A twisted sector in principle gives a product of partition functions if there is more than one orbit, but it will suffice to consider the case of a single orbit. This is because if there are different orbits, the vacuum energy is simply the sum of the vacuum energy for each orbit. In the case of a single orbit, the partition function looks like
\be
Z= \sum_E e^{-\beta E}\,.
\ee
For a generic torus there can be angular potentials, but we have suppressed them since they will not influence the vacuum energy. Note that these values $E$ are not directly the energy on the new spatial torus as there may have been a rescaling of the time direction. The vacuum energy of the orbifold theory $\EV^{\text{orbi}}$ is simply the smallest such value of $E$. Now consider a twisted sector giving an arbitrary rescaling $U_i \to M_i U_i$ such that
\be
\prod_{i=0}^{d-1} M_i = N \label{prodM} \,.
\ee
This is needed as the scaling of the full torus must be equal to $N$ if there is only one orbit. On such a torus, the vacuum contribution will be of the form
\bea\label{rescaledE}
\EV ^{\text{orbi}}(M_i)&=&-M_0\frac{\evac V_{d-1}\prod_{i>0} M_i}{M_1^{d} L_1^{d}}\left(1+f(\mathbf{y}_1)\right)  \notag \\
&=&-\frac{N}{M_1^{d}}\frac{\evac V_{d-1}}{L_1^{d}}\left(1+f(\mathbf{y}_1)\right),
\eea
where we used \rref{prodM} and 
\be
\mathbf{y}_1 = \left(\frac{M_1 L_{1}}{M_2 L_2}, \dots, \frac{M_1L_1}{M_{d-1}L_{d-1}}\right).
\ee
From \rref{rescaledE} and using the monotonicity property of $f(\y)$ under the increase of any of its arguments, it is clear that this expression is maximized for all $M_i=1$ except for $M_1$. At first glance, it is not clear if increasing $M_1$ increases or decreases the energy as it appears both in the denominator and in $f(\y)$ which change in opposite directions. However, one can alternatively write the vacuum energy as
\be
\EV(M_i)=- \frac{N}{M_2^{d}}\frac{ \evac V_{d-1}}{L_2^{d}}\left(1+f(\mathbf{y}_2\right)) \,,
\ee
with
\be
\mathbf{y}_2 = \left(\frac{M_2 L_{2}}{M_1 L_1}, \dots, \frac{M_2L_2}{M_{d-1}L_{d-1}}\right).
\ee
In this form, it is clear that $M_1>1$ would only give a less negative value to the free energy. We have thus showed that to get the minimal contribution, we need
\be
M_0=N, \quad M_i = 1 \ \ \forall\, i\,,
\ee
which then gives precisely the vacuum energy \rref{evacN}.

Although this might appear as good news for the orbifold theory to be a ``nice" theory, it is very bad news for any chance of universality at large $N$. We have shown in the previous section that having $\tf=0$ is a necessary condition for a universal free energy and an extended regime of the Cardy formula. Here, we see that the orbifold theory has $\tf=0$ only if the seed theory does. The choice of seed becomes crucial to reproduce the phase structure of gravity. In fact, this result is not so surprising. In two dimensions, we could consider ourselves lucky that the $S_N$ orbifold theory, which is a free theory, reproduces the phase structure of Einstein gravity. It is only the strong coupling deformation of the orbifold theory that is dual to Einstein gravity so there is no a priori reason why one should have expected the orbifold theory to reproduce the phase structure of gravity. In higher dimensions, it appears that for a general seed, some form of coupling between the $N$ CFTs must be introduced to force $\tf$ to vanish. One might consider deforming the orbifold theory by some operator to achieve this effect. In particular, the existence of any exactly marginal deformations might allow reducing the Hagedorn density of light states to something compatible with Einstein gravity, as is proposed to occur in the D1-D5 duality. This could be directly connected to the vanishing of $\tf$.  

In the following subsection, we will show that choosing a seed theory with $\tf=0$ both gives a theory that saturates the sparseness bound and reproduces the phase structure of gravity.

\subsection{Universality for $\tf=0$ and free energy at large $N$ \label{freeenergyorbi}}
If $\tf=0$, we have $E_\text{vac}=-\varepsilon_\va V_{d-1}/L_1^{d}$ where $L_1$ is the length of the smallest cycle. Inserting this expression in \rref{hagedorngrowth}, we obtain
\be 
\rho(E) \sim \exp\left(L_1(E-N E_\text{vac} )\right) \label{rhofinal}
\ee
for the growth coming from the specific twisted sector we previously considered. Note that the coefficient of the Hagedorn growth precisely saturates the bound on the light states given in \rref{finalbound} if we put the theory on the square torus. At the upper end of the range of validity of \rref{hagedorngrowth} where $E=-(d-1)N\EV$, we precisely recover the Cardy growth at the same energy. This indicates that the spectrum transitions sharply from Hagedorn to Cardy exactly where expected. However, we have only given a lower bound for the density of states as we only computed the contribution coming from a particular twisted sector. We will now show that for $\tf=0$ it is also an upper bound. We will do so by computing the free energy and see that it precisely reproduces the universal behavior discussed in section \ref{proofs}. This implies that the density of low-lying states is bounded above by \rref{rhofinal}, which becomes both a lower and upper bound. This means that no other twisted sector can give a bigger contribution and the density of states is well-approximated by \rref{rhofinal}.

To compute the free energy at large $N$, we will follow a similar procedure as that in two dimensions \cite{Keller:2011xi}. The starting point is a combinatorics formula first introduced by Bantay \cite{Bantay:2000eq}. Let $G$ be a finitely generated group and $Z$ a function on the finite index subgroups of $G$ that takes values in a commutative ring and is constant on conjugacy classes of subgroups.  We have the following identity
\be\label{bantay}
\sum_{N=0}^{\infty} \frac{p^N}{N!} \sum_{\phi:G \to S_N} \prod_{\xi \in \mathcal{O}(\phi)} Z(G_{\xi}) = \exp\left( \sum_{H<G} p^{[G:H]}\frac{Z(H)}{[G:H]} \right),
\ee
where $\phi$ is an homomorphism from $G$ to $S_N$ and $H$ are subgroups of $G$ with finite index given by $[G:H]$. In our case, $Z$ will be the partition function and $G = \pi_1(\mathbb{T}^d) = \mathbb{Z}^d$. This group is abelian and the sum over homomorphisms $\phi$ is equivalent to the sum over commuting elements introduced earlier. The image of $\phi$ acts on $N$ letters (momentarily this will be the $N$ copies of the CFT) by the usual $S_N$ action and its orbit is denoted by $\mathcal{O}(\phi)$. The subgroup $G_{\xi}$ consists of those elements of $g$ such that $\phi(g)$ leaves $\xi$ invariant. In fact, the left hand side is simply the generating function for the partition functions of the symmetric product orbifolds. It corresponds to
\be
\mathcal{Z}=\sum_N p^N Z_N\,,
\ee
where $Z_N$ is the partition function of $\mathcal{C}^{\otimes N}/S_N$ and thus the action of $\phi$ can be thought of as permuting the copies in $\mathcal{C}^{\otimes N}$. Just like in two dimensions, it is often more convenient to work with this generating function and to later find the coefficient of the term $p^N$ to extract $Z_N$.

Bantay's formula equates the generating function to an exponential of a sum over new partition functions. This sum over partition functions really corresponds to a sum over new tori, and for a given index, the volume of the new tori will be the original volume times the index. Just as for $SL(2,\mathbb{Z})$, there is a very natural way to include all tori of a given index by using Hecke operators. Consider a torus to be described by the matrix $U$ given in \rref{Umatrix}, which is upper triangular. Now consider the following set of matrices
\begin{equation}
\Omega_L = \left\{\begin{bmatrix}
a_0 & a_{01}   & \cdots   &a_{0,(d-2)}& a_{0,(d-1)}  \\
0  & a_1    & \cdots &a_{1,(d-2)}& a_{1,(d-1)} \\
\vdots  & \vdots & \ddots   & \vdots& \vdots\\
0  &  0 &\cdots     &a_{d-2} & a_{(d-2),(d-1)}  \\
0  &  0 & \cdots     &0&a_{d-1}
\end{bmatrix}  \Bigg| \prod_i a_i = L, 0\leq a_{j,i} < a_i \ \forall \ i,j
\right\}
\end{equation}
with $L$ fixed. These matrices are elements of $GL(d,\mathbf{Z})$ and act on the lattice vectors $U_i$ defining the torus according to $\widetilde{U} = HU$ with $H$ an element of $\Omega_L$. These new tori will have a volume $L$  times larger than the original torus $U$. Consequently, the new lattice defined by the new torus is a sublattice $H$ of $\mathbb{Z}^d$ and the index $[G:H]$ of $H$ in $G = \mathbb{Z}^d$ is $L$. The purpose of these matrices is to parameterize the finite index subgroups of $G$ so that we can write
\be
\sum_{H<G} p^{[G:H]}\frac{Z(H)}{[G:H]} = \sum_{L>0} \frac{p^L}{L} \sum_{A \in \Omega_L} Z(AU)\,.
\ee
Fortunately, the right hand side can be rewritten in terms of Hecke operators for $SL(d,\mathbb{Z})$, 
\be
T_LZ(U)\equiv\sum_{A\in\Omega_L} Z( A U)\,,
\ee
which encapsulate the sum over different tori mentioned earlier. Note that the Hecke transform of $Z$ is also an $SL(d,\mathbb{Z})$ modular invariant. Bantay's formula then becomes
\be
\mathcal{Z}(U)=\exp\left( \sum_{L>0} \frac{p^L}{L} T_LZ(U)\right). \label{bantay2}
\ee
Because $T_LZ(U)$ is a function invariant under $SL(d,\mathbb{Z})$ \cite{goldfeld2006}, and it has a corresponding extensive free energy, its asymptotic growth is also given by the higher-dimensional Cardy formula. To see this directly, notice that $T_L Z(U)$ is a sum over partition functions of different tori. Each of these obeys the higher-dimensional Cardy formula, although the explicit dependence on the volume of the torus in our higher-dimensional Cardy formula may seem confusing. Note however that at asymptotically large energies we have $E \propto V_{d-1}^{-1/(d-1)}$, so the volume of the torus cancels out and the formula can be written in terms of a dimensionless energy. Thus, there is no confusion as to ``which volume" enters into the Cardy formula for $T_L Z(U)$. In fact, the situation is even better. The gap between the first excited state and the vacuum grows with $L$ indicating that at large $L$, the Cardy formula will become a good estimate for the Hecke transformed partition function.

We are now ready to estimate the free energy. Let us take a rectangular $d$-torus with sides $\beta, L_1, ... , L_{d-1}$, i.e
\be
U=\begin{pmatrix}
\beta & 0   & \cdots   &0& 0  \\
0  & L_1    & \cdots &0& 0 \\
\vdots  & \vdots & \ddots   & \vdots& \vdots\\
0  &  0 &\cdots     &L_{d-2} & 0  \\
0  &  0 & \cdots     &0&L_{d-1}
\end{pmatrix} \,,
\ee
and let us assume $L_1$ is the smallest spatial cycle.
Writing $ \tilde{p} =p e^{\beta \EV}$,
\bea
\mathcal{Z} &=& \exp\left( \sum_{L>0} \frac{\tilde{p}^L}{L} + \sum_{L>0}\frac{\tilde{p}^L}{L}\sum_{E> 0} \widetilde{\rho}_{T_L}(E)e^{-\b E} \right)\notag \\
&=& \left( \sum_{K=0}^{\infty} \tilde{p}^K \right) \exp\left(\sum_{L>0}\frac{\tilde{p}^L}{L}\sum_{E> 0} \widetilde{\rho}_{T_L}(E)e^{-\b E}  \right), \label{generatingfun}
\eea
where we have defined $\tilde{\rho}_{T_L}(E)$ such that
\be
e^{L \b E_\va}T_LZ(U)=1+\sum_{E> 0} \widetilde{\rho}_{T_L}(E)e^{-\b E}
\ee
Using the Cardy formula, the sum over energies in \rref{generatingfun} becomes
\bea
 \sum_{E> 0} 
e^{\left( d a L^{\frac{1}{d}} \left( E + \EV L \right)^{\frac{d-1}{d}}\right)}e^{-\b E}&\sim &\exp\left( L|\EV|\left( \frac{L_1^d}{\b^{d-1}} - \b\right) \right) ,\label{saddleovere}
\eea
where we assumed $L_1$ to be the smallest cycle and used \rref{adef} as well as 
\be
\EV = \frac{-\evac V_{d-1}}{L_1^d} \,.
\ee
The saddle point value for $E$ is
\be
E_s=|\EV| L\left(1 + \frac{(d-1)L_1^d}{\b^d}\right),
\ee
which will be large for large $L$. This justifies the use of the Cardy formula. The terms with low $E$ will of course not be in the Cardy regime but these will only give a subleading contribution. Overall, the error on the each term in the sum over $L$ will be of order $e^{-u L /\beta^{d-1}}$ for some positive order one number $u$ that is theory dependent. Plugging \rref{saddleovere} into \eqref{generatingfun} we get
\bea
\mathcal{Z}&=&\left( \sum_{K=0}^{\infty} \tilde{p}^K \right) \exp\left(\sum_{L>0}\frac{1}{L}\left(\tilde{p}\exp\left( |\EV|\left( \frac{L_1^d}{\b^{d-1}} - \b\right) \right) \right)^L \right) \notag \\
&=&\left( \sum_{K=0}^{\infty} \tilde{p}^K \right) \exp\left( - \log\left(1-\tilde{p}e^{|\EV|\b(L_1^d/\b^d - 1)} \right)  \right) \notag \\
&=&\left( \sum_{K=0}^{\infty} \tilde{p}^K \right)\frac{1}{1-\tilde{p}e^{|\EV|\b(L_1^d/\b^d - 1)}}\,. \label{generatingfinal}
\eea
We can now extract the free energy. Note that because the vacuum energy is negative and proportional to $N$, the partition function diverges as $N\to\infty$ so we need to consider the shifted partition function and shifted free energy
\bea
\widetilde{Z}&\equiv& e^{\EV \b}Z \,,\notag \\
\widetilde{F}&\equiv&-\frac{ \log \widetilde{Z}}{\b}\,. \,.
\eea
The shifted partition function will then simply be the term $\tilde{p}^N$ in \rref{generatingfinal}, which is given by
\be
\widetilde{Z}_N = \frac{\exp\left((N+1)|\EV|\b\left( \frac{L_1^d}{\b^d} - 1\right)\right)-1}{\exp\left(|\EV|\b\left( \frac{L_1^d}{\b^d} - 1\right)\right)-1}\,.
\ee
The free energy as  $N\to\infty$ for $\b < L_1$ is thus
\be
\widetilde{F}_N(U) = -N|\EV|\left( \frac{L_1^d}{\b^d} - 1 \right) \,.
\ee
For $\b > L_1$, we get
\be
\widetilde{F}_N(U) = \frac{1}{\b}\log\left(1 - \exp\left(|\EV|\b\left( \frac{L_1^d}{\b^d} - 1\right)\right)  \right)  + F_{\text{cor}}(\beta)\,,
\ee
where the $F_{\text{cor}}(\beta)$ corresponds to another $\mathcal{O}(1)$ contribution coming from subleading corrections to the saddle point as well as the low energy contributions. The free energy thus has a phase transition at $\b=L_1$ and goes from being $\mathcal{O}(1)$ to $\mathcal{O}(N)$. This precisely matches the phase structure of the bulk gravitational theory.

Modular invariance is not manifest in the shifted free energy above. In order to recover it, we consider the quantity
\be
\mathscr{F}(U) = \lim_{N\to \infty} \frac{1}{N}F_N(U)\,,
\ee
where $F_N(U)$ is the unshifted free energy and $\mathscr{F}(U) = \mathscr{F}(\b,L_1,...,L_{d-1})$. Using the results obtained above,
\be\label{largeNfreeenergy}
\mathscr{F}(U)  = \left\{
\begin{array}{ll}
-\frac{\evac V_{d-1}}{\b^d} & \b < L_1\\[0.2cm]
-\frac{\evac V_{d-1}}{L_1^d} & \b > L_1
\end{array}
\right.,
\ee
where $L_1$ is the smallest cycle. The free energy is a modular covariant quantity which transforms under the $S$ transform of $SL(d,\mathbb{Z})$ as 
\be
\mathscr{F}(\b,L_1, ... ,L_{d-1}) = \frac{L_1}{\b}\mathscr{F}(L_1, ... , L_{d-1},\b).
\ee
Upon checking this transformation rule for \rref{largeNfreeenergy}, we see that in both regimes the free energy transforms as expected.

\section{Discussion}\label{disc}


In this paper we have studied conformal field theories in dimensions $d>2$ compactified on tori. The main
goal was to explore the implications of the assumed invariance under the 
$SL(d,\mathbb Z)$ modular group and see what additional constraints on the spectrum would reproduce the phase diagram of gravity in anti-de Sitter space. We have uncovered both similarities and differences with the two-dimensional case. We have presumably only scratched the surface of this interesting subject and many issues and open questions remain, some of which we list below.

\subsection{Modular invariance}

The modular group $SL(d,\mathbb Z)$ consists of the large diffeomorphisms (i.e. not continuously connected
to the identity element) which map a $d$-dimensional torus to itself. In two dimensions, there are well-known
systems, such as the chiral fermion, whose partition function is not modular invariant. However, such theories
have gravitational anomalies and can therefore a priori not be consistently defined on arbitrary manifolds. Moreover, 
when such theories appear in nature, as in the edge modes in the quantum Hall effect, the relevant anomalies are
canceled due to an anomaly inflow mechanism which crucially relies on the existence of a higher-dimensional system 
to which the theory is coupled (for a higher-dimensional version of this statement see e.g. \cite{Park:2016xhc}).
We are not aware of a local and unitary conformal field theory which is free of local gravitational
anomalies and not modular invariant. But modular invariance is weaker than the absence of local gravitational
anomalies. There are many modular invariant CFTs with $c_L-c_R\neq 0$ which have gravitational anomalies, while modular
invariance only implies that $c_L-c_R$ must be an integer multiple of 24. It would be interesting to explore
the generalizations of these statements to higher dimensions.

Another approach to using modular invariance to learn about conformal field theories on tori is to consider bounds coming from the fixed points of $SL(d,\mathbb{Z})$. This would be a generalization of the ``modular bootstrap" \cite{Cardy:1991kr,  Hellerman:2009bu, Hellerman:2010qd, Friedan:2013cba,  Qualls:2013eha, Qualls:2014oea, Benjamin:2016fhe,  Collier:2016cls} to higher dimensions. This is valid for general conformal field theories, and taking a large-$N$ limit may give insight into holographic theories. 

\subsection{State-operator correspondence}

The usual arguments for the state-operator correspondence in conformal field theory rely on radial quantization and
apply to the theory on the spatial sphere $S^{d-1}$ times time. The local operators obtained
in this way can be inserted on other manifolds as well but the one-to-one correspondence with states in the Hilbert space
no longer applies. The main problem in applying radial quantization to the torus is that, as opposed to spheres,
one can not smoothly shrink a torus of dimension larger than one to a point. Stated more precisely, the metric
$ds^2=dr^2+r^2 d\Omega^2$ is not smooth at $r=0$ unless $\Omega$ is the round unit sphere.

One cannot even apply the standard radial quantization argument to the conformal field theory on $S^1 \times {\mathbb R}^{d-2}$ times time. At $r=0$, the metric $ds^2=dr^2+r^2d\phi^2+r^2 dx_i dx_i$ looks like a singular
${\mathbb R}^{d-2}$-dimensional plane, suggesting that some sort of surface operators might be relevant.
That such operators are
generically needed can for example be seen using the orbifold theories we studied in this paper. Orbifold theories can
be thought of as theories with a discrete gauge symmetry, and in case the theory lives on $S^1 \times {\mathbb R}^{d-2}$
we should include twisted sectors which involve twisted boundary conditions when going around the $S^1$. These twisted
boundary conditions can be detected by a Wilson line operator for the discrete gauge field around the $S^1$. To create
a non-trivial expectation value for the Wilson line operator, we need an operator which creates non-contractible loops, and for this we need an operator localized along a $(d-2)$-dimensional surface. One can think of such operators as a higher-dimensional generalization of the 't Hooft line operators. A local operator in $d>2$ is unable to generate a non-trivial
vev for the discrete Wilson line operator and can therefore not create twisted sector states. Surface operators of
dimension $d-2$ which create twisted boundary conditions also feature prominently in the replica trick
computations of entanglement entropy in dimensions $d>2$; they are the generalized twist fields associated to the
boundary of the entangling area.

If $(d-2)$-dimensional surface operators are the right operators for the theory on $S^1 \times {\mathbb R}^{d-2}$,
it is plausible they are also relevant for CFT's on tori. One can for example consider the surface operators dual
to periodic field  configurations on ${\mathbb R}^{d-2}$,
but it is not clear the resulting surface operator will have the right periodicity as well.
Alternatively, one can study the Euclidean theory on an annulus times $\mathbb{T}^{d-2}$, with the annulus having inner radius $R_1$
and outer radius $R_2$. The Euclidean path integral in principle provides a map from states on the torus $S^1_{R_1}\times \mathbb{T}^{d-2}$
to $S^1_{R_2}\times \mathbb{T}^{d-2}$, and by taking the limit $R_1\rightarrow 0$ one can imagine obtaining singular boundary conditions
for a surface operator localized along a $(d-2)$-torus.

Clearly, more work is required to understand whether the above construction provides a useful version of the state-operator
correspondence for field theories on tori, and if it does, what a useful basis for the space of surface operators could possibly be.
There seems to be a significant overcounting, as one can construct a surface operator for any choice of state on the torus and
for any choice of one-cycle on the torus. Currently, we do not even have a compelling compact Euclidean path integral representation of 
the ground state of the theory on the torus.

It might also be interesting to explore the state-operator correspondence from an AdS/CFT point of view. One would then need
to glue Euclidean caps to the Lorentzian solutions discussed in section \ref{phases}. Since the Lorentzian solutions
require a choice of one-cycle which is smoothly being contracted in the interior, a similar choice will be needed for the Euclidean
caps, leading apparently once more to the same overcounting as we observed above. It would still be interesting to construct the 
explicit form of the geometry where a Euclidean cap without the insertion of surface operators 
is smoothly glued to the Lorentzian AdS solutions. If such solutions could be found, its boundary geometry
would provide a Euclidean path integral description of the ground state of the corresponding CFT, at least in the large $N$ and 
strong coupling approximation. 

\subsection{Defining the orbifold theory}

In section \ref{section:orbifolds}, we defined a prescription to compute the partition function of the orbifold theory. This prescription describes both the Hilbert space and the spectrum of the Hamiltonian on the torus. In two dimensions, the orbifolding prescription also fully describes the procedure to compute arbitrary correlation functions of (un)twisted sector local operators, at least in principle. In higher dimensions, because of the lack of a precise state-operator correspondence, it is not clear wether we have really fully specified a theory. For that, we need to determine the full set of correlation functions and hence know the set of operators in the theory. It is clear that all untwisted sector correlation functions make sense in the orbifold theory so all local correlation functions are well-defined and calculable. Furthermore, the theory possesses a stress tensor as the stress tensor is always in the untwisted sector. Nevertheless, the questions touching the twisted sector states and/or line operators is much more obscure and it would be very interesting to understand the extent to which the orbifolding prescription fully determines these.

One way orbifold theories in higher dimensions can potentially appear (and therefore inherit a natural definition) are as discrete gauge theories that arise in the infrared limit of a gauge theory with spontaneously broken continuous gauge symmetry (e.g. $SU(N)\rightarrow S_N$). This would also explain how to couple the theory to other manifolds, an issue we turn to in the next section.

\subsection{Orbifold theories on other manifolds}

The orbifold theories we studied are most easily defined on tori. However, if we have fully defined a theory we should be able to put it on any manifold. Viewing them as theories with a discrete gauge group
also provides a prescription for the sum over twisted sectors when computing the path integral for other manifolds.
The sum over twisted sectors is the same as the sum over the space of flat connections modulo an overall conjugation, and for a manifold $M$ this space is ${\rm Hom}(\pi_1(M),G)/G$. But even for flat space, where no sum over twisted sectors needs to be taken, there are still signs of the discrete gauge symmetry. In particular, one can consider surface operators which create twisted sector states even on the plane, and their correlation functions contain interesting new information. Such operators naturally arise in the context of Renyi entropy calculation in higher dimensions \cite{Hung:2011nu, Hung:2014npa}.

\subsection{Outlook}

We have only begun to explore the properties of modular-invariant field theories on tori and their role in AdS/CFT. The interesting relations between the form of the ground state energy, universal free energy at high-temperature, sparseness conditions on the spectrum and vacuum dominance in the partition function beg for a deeper understanding. Is there a more precise relation between the low- and high-energy spectrum that can be rigorously established? Can subleading corrections be systematically analyzed? How much of the rich structure in $d=2$ and the mathematics of $SL(2,\mathbb Z)$ can be carried over to $d>2$? Does all this shed any new light on which theories can have weakly coupled gravitational duals? 

\section*{Acknowledgements}
The authors would like to acknowledge helpful conversations with Tom Hartman, Diego Hofman, Christoph Keller, and Alex Maloney. AB is supported by the Foundation for  Fundamental Research on Matter (FOM). JK is supported by the Delta ITP consortium, a program of the Netherlands Organisation for Scientific Research (NWO) that is funded by the Dutch Ministry of Education, Culture and Science (OCW). BM and ES are supported by NSF Grant PHY13-16748. MS is supported by the Yzurdiaga Chair at UCSB and the Simons Qubit Collaboration. MS would like to thank the Stanford Institute of Theoretical Physics for hospitality during part of the completion of this work.

\appendix

\section{Alternate proof without assuming $\tf= 0$}\label{alternate}
In this section we will try to generalize the proof in section \ref{fneq0} to the case where we make no assumptions on the form of the vacuum energy. To illustrate the point, we will work in three spacetime dimensions and make the spatial manifold explicit. We will again be using a proof like that of \cite{Hartman:2014oaa}, but this time we will take $N\rightarrow \infty$ from the start.\footnote{Thanks to Tom Hartman for discussions about this simpler form of proof.}

Consider a rectangular three-torus with side lengths $\beta$, $L_1$, and $L_2$ with   $\beta<L_1<L_2$. We have the relations
\begin{align}
Z(\beta)_{L_1\times L_2} - Z(L_1)_{\beta \times L_2}&=Z(\beta)_{L_1\times L_2} - Z(L_2)_{\beta \times L_1}=0
\implies\\
\left(\sum_L e^{-\beta E_{L_1\times L_2}}-\sum_L e^{-L_1 E_{\beta \times L_2}}\right)&+\left(\sum_H e^{-\beta E_{L_1\times L_2}}-\sum_H e^{-L_1 E_{\beta \times L_2}}\right)=0\,,\\
\left(\sum_L e^{-\beta E_{L_1\times L_2}}-\sum_L e^{-L_2 E_{\beta \times L_1}}\right)&+\left(\sum_H e^{-\beta E_{L_1\times L_2}}-\sum_H e^{-L_2 E_{\beta \times L_1}}\right)=0\,.
\end{align}
Notice that light states $L$ and heavy states $H$ are playing triple duty, since the spatial background changes in the different quantizations. In any given quantization, the states $L$ refer to negative energy states that scale with a positive power of $N$ while $H$ refers to positive energy states that scale with a positive power of $N$. We eliminate the consideration of states with $\mathcal{O}(1)$ energies by bounding their density of states so that their contribution is $\mathcal{O}(1)$ and therefore subleading.

We now assume that for $\beta < L_1$ and $\beta < L_2$, we have 
\begin{align}
\label{toprove0}
\sum_L e^{-L_1 E_{\beta \times L_2}} \gg \sum_L e^{-\beta E_{L_1 \times L_2}}, \qquad
\sum_H e^{-L_1 E_{\beta \times L_2}} \ll \sum_H e^{-\beta E_{L_1 \times L_2}},\\ \sum_L e^{-L_2 E_{\beta \times L_1}} \gg \sum_L e^{-\beta E_{L_1 \times L_2}}, \qquad
\sum_H e^{-L_2 E_{\beta \times L_1}} \ll \sum_H e^{-\beta E_{L_1 \times L_2}}\,.\label{toprove}
\end{align}
These inequalities can be proven to be true in two spacetime dimensions and for the special torus in a general number of dimensions. In fact, it is what makes a proof like that of \cite{Hartman:2014oaa} work. 

Using these inequalities, we can approximate the above equalities as 
\be
\sum_L e^{-L_1 E_{\beta \times L_2}} \approx \sum_H e^{-\beta E_{L_1\times L_2}},\qquad \sum_L e^{-L_2 E_{\beta \times L_1}} \approx \sum_H e^{-\beta E_{L_1\times L_2}}.
\ee
Then we can use $Z_H(L_1)_{\beta \times L_2}\ll Z_H(\beta)_{L_1\times L_2} \approx Z_L(L_1)_{\beta \times L_2}$ and $Z_H(L_2)_{\beta \times L_1}\ll Z_H(\beta)_{L_1\times L_2} \approx Z_L(L_2)_{\beta \times L_1}$ to approximate the partition function in the $L_1$ and $L_2$ channels as  
\begin{align}
Z(L_1)_{\beta \times L_2} &= Z_L(L_1)_{\beta \times L_2}+Z_H(L_1)_{\beta \times L_2} \approx Z_L(L_1)_{\beta \times L_2}\,,\\
Z(L_2)_{\beta \times L_1} &= Z_L(L_2)_{\beta \times L_1}+Z_H(L_2)_{\beta \times L_1} \approx Z_L(L_2)_{\beta \times L_1}\,.
\end{align}
We see that under the assumptions \rref{toprove0} and \rref{toprove}, the partition function is vacuum dominated in the $L_1$ and $L_2$ channels if and only if 
\be
\rho(E_{L_1 \times L_2}<0)\lesssim e^{L_1 (E-E_\va)_{L_1 \times L_2}}.\label{formal}
\ee
As explained in section \ref{proofs} this is necessary and sufficient for a universal free energy at all temperatures on an arbitrary spatial torus.

In general dimension, the sufficient conditions for a universal free energy are the $d-1$ inequalities that generalize \eqref{toprove} and a sparse light spectrum:
\be
\rho(\D) \lesssim \exp\left(L_{\text{min}} \D\right)\,,
\ee
where $L_{\text{min}}$ is the minimum cycle size of the spatial torus.

\section{Microcanonical density of states}\label{microdense}
The results derived in the main text are phrased in terms of the canonical partition function $Z(\beta)$. In general such results do not immediately translate into statements about the microcanonical density of states. However, as discussed carefully for two dimensions in \cite{Hartman:2014oaa}, the limit $N\rightarrow \infty$ is a good thermodynamic limit which allows us to conclude $\rho(\langle E\rangle) \approx e^{S(\langle E\rangle)}$ for $\langle E\rangle = -\p_\beta \log Z(\beta)$. Large $N$ suppresses the fluctuations in $\langle E \rangle$ and unambiguously defines an energy $E\equiv \langle E \rangle$. The arguments of \cite{Hartman:2014oaa} carry over straightforwardly and imply that the Cardy density of states has an extended range of validity that holds down to $E=-(d-1)E_\va$, which is the energy corresponding to $\beta = L_1$. Instead of repeating those arguments we will give an alternative way of understanding the thermodynamic limit which gives additional intuition. In the next sections we will evaluate the inverse Laplace transform connecting the canonical partition function to the microcanonical density of states in several examples. The existence of a stable saddle point is the statement of an equivalence between the two ensembles.

In section \ref{b1} we will consider the case of two-dimensional CFT and use the fact that the partition function is dominated by the light states \cite{Hartman:2014oaa}. In section \ref{b2} we will assume $\tf=0$, in which case the special torus allows us to get away with only bounding the density of light states, just as in two dimensions and as we saw in the main text. Finally, in section \ref{b3} we will consider an extension to angular momentum. Here, to extend the regime of validity we will have to bound the density of states for the entire spectrum, again as we saw in the main text.
\subsection{$d=2$}\label{b1}
We will begin by considering the case of two-dimensional conformal field theories, treated in \cite{Hartman:2014oaa}. Here we will directly evaluate the inverse Laplace transform connecting the canonical partition function to the density of states.

We begin with the expression for the degeneracy

\begin{equation}
\rho(h_s,\bar{h}_s) = \int_{i\alpha-\infty}^{i\alpha+\infty} d\tau \int_{-i\alpha-\infty}^{-i\alpha+\infty}   d\bar{\tau}\, I(\tau, \bar{\tau})\,\tilde{Z}(-1/\tau,-1/\bar{\tau})\,,
\end{equation}
for $\alpha>0$, where
\begin{equation}
I(\tau,\bar{\tau}) \equiv e^{-2\pi i \tau (h_s-c/24)}e^{2\pi i \bar{\tau}(\bar{h}_s-\bar{c}/24)} \times \;e^{-2\pi i/\tau(-c/24)}e^{2\pi i/\bar{\tau}(-\bar{c}/24)}~.
\end{equation}
Evaluating the integral using the saddle point approximation for large $h_s$ and $\bar{h}_s$ requires solving the saddle equations $I^{(1,0)}(\tau, \bar{\tau})=I^{(0,1)}(\tau, \bar{\tau})=0$, which gives the dominant saddle
\begin{align}\label{saddles}
\tau_s&=\tau_1^s+i\tau_2^s=+i\sqrt{\frac{c}{24h_s-c}}\;,\\
\bar{\tau}_s&=\tau_1^s-i\tau_2^s=-i\sqrt{\frac{\bar{c}}{24\bar{h}_s-\bar{c}}}\,.\;
\end{align}
Evaluating the integrand on this saddle gives the entropy
\be\label{cardyformula}
S=\log \rho(h_s,\bar{h}_s)= 2\pi\left(\sqrt{\frac{c}{6}\left(h_s-\frac{c}{24}\right)}+\sqrt{\frac{\bar{c}}{6}\left(\bar{h}_s-\frac{\bar{c}}{24}\right)}\right).
\ee
To ensure the saddle point approximation is justified, we have to check that $\tilde{Z}$ does not make big contributions on the saddle:
\begin{align}
\tilde{Z}(-1/\tau_s,-1/\bar{\tau}_s)=\sum_{h,\bar{h}} \rho(h,\bar{h})\,\exp\left(\frac{-2\pi h}{\sqrt{\frac{c}{24h_s-c}}}-\frac{2\pi \bar{h}}{\sqrt{\frac{\bar{c}}{24\bar{h}_s-\bar{c}}}}\right).
\end{align}
As $h_s\rightarrow \infty$ and $\bar{h}_s\rightarrow \infty$ with $c$ finite, all terms except for the vacuum contribution are infinitely exponentially suppressed, justifying our saddle point approximation. This is the ordinary Cardy formula. 

Now let us consider the limit $c\rightarrow \infty$ with $h_s =m c$. For simplicity we set $c=\bar{c}$ and $\tau_1=0$. So we have the canonical partition function at inverse temperature $\beta = \tau_2/(2\pi)$. We will have the same saddle as before but need to check again that $\tilde{Z}$ does not give a big contribution on the saddle. If we take $m\rightarrow \infty$, then again all terms except the vacuum contribution are infinitely exponentially suppressed and our saddle is justified. But now we want to see how small we can make $m$. We will use the fact that $Z(\beta)$ is dominated by the light states as long as $\beta > 2\pi$. This means that $\tilde{Z}(\beta)$ is also dominated by the light states. We can therefore write
\be
\widetilde{Z}(4\pi^2/\beta_s) \approx \sum_{\Delta\leq c/12+\epsilon} \rho(\Delta) \exp\left(\f{2\pi \Delta}{\sqrt{\f{c}{12\Delta_s-c}}}\right)
\ee
for $\D = h+\bar{h}$. We need all terms on the right-hand-side to contribute exponential suppressions, except for the identity operator which will contribute $+1$. To push the validity of the saddle down to $\Delta_s = c/6$, which is the result expected from gravity, we need to bound the degeneracy as 
\be
\rho(\Delta\leq c/12+\epsilon)\lesssim\exp\left(2\pi\Delta\right)=\exp\left(2\pi(E+c/12)\right).
\ee
This is the same bound on the light states as in \cite{Hartman:2014oaa}.

\subsection{$d>2$}\label{b2}
In this case, we consider the high-temperature/low-temperature duality on the special torus, for which $Z(\beta) = Z(L^d/\beta^{d-1})$. We have
\be
\rho(E_s) = \frac{1}{2\pi i}\int_{\a - i\infty}^{\a + i\infty} d\beta \,Z(\beta)\,e^{\beta E_s}
\ee
for $\a>0$. Performing a modular transformation and multiplying and dividing by a common factor gives (omitting the integration limits and $1/2\pi i$)
\be
\rho(E_s) = \int d\beta \,\left(e^{- \varepsilon_\va V_{d-1} /\beta^{d-1}}Z(L^d/\beta^{d-1})\right) e^{\varepsilon_\va V_{d-1} /\beta^{d-1} + \beta E_s}\,.
\ee
We will hold off on defining $V_{d-1}$ for the moment, which will actually be defined to be independent of $\beta$. At large $N$ with $E_s$ scaling as a positive power of $N$ the saddle point (ignoring the term in parentheses) occurs at 
\be
\beta_s = \left(\f{(d-1) \varepsilon_\va V_{d-1}}{E_s}\right)^{\f{1}{d}},
\ee
which gives an on-shell entropy of 
\be
\rho(E_s) =\exp\left( \f{d}{(d-1)^{\f{d-1}{d}}}(\varepsilon_\va V_{d-1})^{\f{1}{d}}E_s^{\f{d-1}{d}}\right)
\ee
To make sure the saddle is controlled, we again want the term in parentheses to contribute as  $1 + e^{-(\dots)}$ on the saddle. To show this, we will use the fact from the main text that  for $Z(L^d/\beta^{d-1})$ is dominated by the contribution of the light states for $L>\beta$. This lets us write the term in parentheses on the saddle as
\be
 \sum_{E<\epsilon} \exp\left(-(L^d E+\evac V_{d-1}) \left(\f{E_s}{(d-1) \varepsilon_\va V_{d-1}}\right)^{\f{d-1}{d}}\right)\,.
\ee
We now want to define $V_{d-1}=V_\Mz =L\cdots L^{d-1}/\beta_s^{d-2}$ as the volume of the special torus. Since $\beta_s$ depends on $V_{d-1}$, this is an equation that can easily be solved for $V_{d-1}$ in terms of only $\varepsilon_\va$ and $E_s$, but all we need to know is that it gives the volume of the special torus on the saddle. Now using our assumption that the subextensive corrections to the vacuum energy vanish, we see that the vacuum state contributes as $+1$. To approach the square torus as in the main text we want to push $E_s$ down to $-(d-1)E_\va$, which will require bounding the density of light states as 
\be
\rho(E) \lesssim \exp(L(E-E_\va)), \qquad E \leq -(d-1)\, E_\va\,,
\ee
where the energies are taken to be on a square torus. This is the same bound as we saw in the main text. At this point we can perform a similar bootstrapping procedure to obtain this density of states on an arbitrary spatial torus and at arbitrarily higher energies.

\subsection{Cardy extension with angular momentum on $\mathbb{T}^2 \times \mathbb{R}^{d-2}$}\label{b3}
We will show in this section that similar manipulations can be performed once angular momentum is included. In particular, assuming sparseness on the low-lying spectrum, we can extend the generalized Cardy formula with angular momentum to include the entire range
\be
J^2<(E-E_\va)(E+(d-1)E_\va)\,.
\ee
Note that this has the correct limits. For $d=2$ we recover $E_L E_R>c^2/576$, and for $J=0$, $d>2$ we get $E>-(d-1) E_\va$. 


Before we perform our CFT analysis, we should analyze the phase structure of gravity with the appropriate boundary conditions. We are introducing a chemical potential for angular momentum, which corresponds to adding a twist in the periodicity of Euclidean time.  The solutions are the same as in the main text, but with angular velocity added. The Poincar\'e patch and soliton geometries can be written as before except with the new identification $t\sim t+i\beta+\theta$, while the black brane is written as 
\begin{align}
ds^2&=\left((r_h/r)^{d}u_\mu u_\nu + r^2 \eta_{\mu\nu}\right)dx^\mu dx^\nu + \f{dr^2}{r^2\left(1-(r_h/r)^{d}\right)}\,,\\
 u_\mu &= \left(\f{-1}{\sqrt{1-a^2}}\,,\f{a}{\sqrt{1-a^2}},\vec{0}\right)\,.
\end{align}

 The free energies of the solutions are given by
\be
F_{\textrm{bb}}= -\frac{r_h^{d} L L_{\infty}^{d-2} }{16\pi G }\,, \qquad   F_{\textrm{sol}}= -\frac{r_0^{d} L L_{\infty}^{d-2} }{16\pi G }\,, \qquad F_{\textrm{pp}} = 0\,,
\ee
with
\be
r_h= \frac{4\pi}{d}\sqrt{\frac{1}{\beta^2 + \theta^2}}\,,  \qquad r_0= \frac{4\pi}{d L}\,.
\ee
The energy and the angular momentum of the black brane are given by the usual thermodynamic relations in terms of a Euclidean partition function $\mathcal Z(\beta,\theta) = \Sigma \ e^{-\beta H + i \theta J} =  \Sigma \ e^{-\beta(H + a J)}$: 
\begin{align}
E  &=-\frac{\partial}{\partial\beta} \bigg |_\theta \ \mathrm{log}  Z = \frac{r_h^{d}LL_\infty^{d-2} }{16\pi G }\frac{d-1+a^2}{1-a^2}\,, \\
 J &=-i \frac{\partial}{\partial \theta} \bigg |_\beta   \ \mathrm{log}  Z =  \frac{r_h^{d}LL_\infty^{d-2} }{16\pi G } \frac{da}{(1-a^2)} \,.
\end{align}
From the expressions for the free energies, we see that the soliton dominates the ensemble for $r_0>r_h$. At the phase transition $r_h= \frac{4\pi}{dL}$, the energy and angular momentum are related by 
\be\label{rep}
J^2=(E-E_\va)(E+(d-1)E_\va)\,.
\ee
We now turn to our CFT analysis. The canonical partition function at finite temperature and angular velocity is defined as 
\begin{align*}
& Z(\tau,\bar\tau) = \mathrm{Tr}\left(e^{2\pi i \tau E_R} e^{-2\pi i \bar\tau E_L}\right), \\
 E_R & +  E_L = E\,, \qquad  E_R - E_L = J\,,
\end{align*}
where $\tau=re^{i\phi}$ is the modular parameter whose imaginary part acts as the inverse temperature, and the real part acts as the chemical potential for angular momentum. We have only turned on a single angular momentum generator. The microcanonical density of states is given by the usual inverse Laplace transform (up to subleading Jacobian factors which we ignore): 
\begin{equation}
\rho(E_s,J_s) =\int dr d\phi \  Z(r,\phi) \ \mathrm{exp}\left[ - \pi i r e^{i\phi}(E_s+J_s) +  \pi i r e^{-i\phi}(E_s-J_s)  \right].
\end{equation}
For simplicity, we will work in the special case of $\mathbb{T}^2 \times \mathbb{T}^{d-2}_\infty$ and consider the angular momentum to be along the spatial cycle of the $\mathbb{T}^2$. On this background, modular invariance gives 
\be
\text{log}\  Z(r,\phi) \approx  r^{2-d} \text{log}\ Z(-r^{-1},-\phi)\,.
\ee
As before we define a shifted partition function as
\begin{equation}
\widetilde{Z}(r,\phi) \equiv  \mathrm{Tr}\;\exp\left[ \pi i r e^{i\phi}(E_s+J_s-E_\va) -  \pi i r e^{-i\phi}(E_s-J_s-E_\va)   \right].
\end{equation}
Using the above we write the density of states as
\begin{align}
\rho(E_s,J_s) = \int dr \ d\phi \  & \widetilde{Z}\left(-\frac{1}{r},-\phi \right)^{r^{2-d}}  \mathrm{exp}\left[ \frac{-\pi i E_\va}{r^{d-1} e^{i\phi}} + \frac{\pi i E_\va}{r^{d-1} e^{-i\phi}} \right] \nonumber \\
& \qquad \ \ \times \mathrm{exp} \left[ - \pi i r e^{i\phi}(E_s+J_s) + \pi i r e^{-i\phi}(E_s-J_s)  \right].
\end{align}
At large $N$, we can approximate the above integral by its saddle-point value, which gives
\begin{align}
\rho(E_s,J_s) & = \mathrm{exp} \bigg[ \pi  \sqrt{d} \left(\frac{2}{(d-1)^{d-1}}\right)^{\frac{1}{d}}  \left(\sqrt{d^2 E_s^2 - 4(d-1)J_s^2}-(d-2) E_s \right)^{\frac{d-2}{2 d}} \nonumber \\
& \qquad \qquad \qquad \quad \times \left(\sqrt{d^2 E_s^2-4 (d-1) J_s^2}+d E_s \right)^{1/2}(-E_\va)^{\frac{1}{d}} \ \bigg].
\end{align}
This is the higher-dimensional Cardy formula with angular momentum. To ensure that our saddle is controlled and this formula is valid, we need to check that the neglected piece $\widetilde Z$ is not large on the saddle. By definition
\begin{align}
\widetilde{Z} \left(-\frac{1}{r_s},-\phi_s \right) = 
  \int\displaylimits_{\textrm{light}} dE dJ \ \rho(E,J) \ \mathrm{exp} \left[-\frac{\pi i  }{r_s}\left(e^{-i\phi_s}\left(\D+J\right) - e^{i\phi_s} \left(\D - J\right)\right)  \right] \nonumber \\
\hspace{-8mm}+ \int\displaylimits_{\textrm{heavy}}  dE dJ  \ \rho_{\text{\tiny{Cardy}}} (E,J) \ \mathrm{exp} \left[-\frac{\pi i  }{r_s}\left(e^{-i\phi_s}\left(\D+J\right) - e^{i\phi_s} \left(\D - J\right)\right)  \right],
\end{align}
where we have used $\D = E-E_\va$. We would like to find and maximize the range in the spectrum where the heavy states lie. The first line stands for the contribution of light states and is $\mathcal{O}(1)$ as long as the density of light states obeys a Hagedorn bound. The second line is small if
\begin{equation}
\log \rho_{\text{\tiny{Cardy}}} - \frac{\pi i  }{r_s} \left( e^{-i\phi_s} \left(\D+J \right) - e^{i\phi_s} \left(\D - J \right) \right) < 0\,. 
\end{equation}
Let us denote the left hand side of this expression by $\mathcal{T}(E_s, J_s, E, J)$. The dependence of $\mathcal{T}$ on $E_s$ and $J_s$ comes through $r_s$ and $\phi_s$. Using the values of the saddle and the Cardy formula gives a messy expression for $\mathcal{T}(E_s, J_s, E, J)$.

We would like to find the region in the $E,J$ plane where $\mathcal{T}(E_s,J_s,E,J)<0$. Note that since $\mathcal{T}(E_s, J_s, E, J)$ is also a function of $E_s$ and $J_s$, this region will depend on the values of $E_s$ and $J_s$. This means we need to find the values of $E_s$ and $J_s$ for which the region in the $E,J$ plane is maximized. To guarantee that $\mathcal{T}$ is less than zero in a given region in the $E,J$ plane, it will be sufficient to show that the maximum value of $\mathcal{T}$ with respect to $E$ and $J$ is smaller than zero in that region. Saturating this bound will give us the extended range of validity of the Cardy formula. In other words, maximization of $\mathcal{T}$ with respect to $E$ and $J$ will give us $E$ and $J$ in terms of $E_s$ and $J_s$. Then demanding the maximum of $\mathcal{T}$ to be smaller than zero will give a constraint on how small we can make $E_s$ and $J_s$. 

Let's see this in the simpler case of $d=2$ and $J=0$:
\begin{equation}
\mathcal{T}(\D,\D_s) = 2\pi \sqrt{\frac c 3 \left( \D - \frac{c}{12} \right)} - \frac{2\pi \D}{\sqrt{\frac{c}{12\D_s-c}}} \,.
\end{equation}
Extremizing with respect to $\D$ gives 
\begin{equation}
\D_\star=\frac{c \D_s}{12\D_s-c}\,, \qquad  \mathcal{T}_{max} = \f{\pi}{3}(c-6\D_s) \sqrt{\frac{c}{12 \D_s - c}}\,.
\end{equation}
Imposing $\mathcal{T}_{max} \le 0$ gives
\begin{equation}
\D_s \ge c/6\,.
\end{equation}
Hence, we find that using this method we can safely extend the validity of the Cardy formula to the range $\D_s \ge c/6$. For energies smaller than that $\widetilde{Z}$ stops being $\mathcal{O}(1)$ and the saddle point analysis is not valid. Note that in this method the contribution of light states $(\D<c/6)$ was made $\mathcal{O}(1)$ by imposing a Hagedorn bound $\rho(\D<c/6)\lesssim \exp(2\pi \D)$. Here we have not used the result from \cite{Hartman:2014oaa} that the partition function at large $N$ is dominated by the states with $\D\lesssim c/12$, which would allow us to only place a Hagedorn bound on those states.
Proceeding similarly for arbitrary $d$ and nonzero $J$, we find Cardy behavior for the range
\begin{equation}
(d-1) E E_\va-d E_\va^2+E^2>J^2\,,
\end{equation}
which is identical to the bulk result \eqref{rep}.

\small{
\bibliographystyle{JHEP}
\bibliography{cardyextendref}
}

\end{document}